\documentclass[
reprint,
superscriptaddress,
preprintnumbers,
 amsmath,amssymb,
 aps,showpacs,showkeys
]{revtex4-1}

\usepackage{graphicx}
\usepackage{dcolumn}
\usepackage{bm}
\begin{document}

\title{Observation of bifurcations and hysteresis in experimentally coupled logistic maps}

\author{Carac{\'e} Guti{\'e}rrez}
\affiliation{Universidad de la Rep{\'u}blica, Instituto de F{\'i}sica de Facultad de Ciencias, Igu{\'a} 4225, Montevideo 11400, Uruguay.}
\author{Cecilia Cabeza}
\affiliation{Universidad de la Rep{\'u}blica, Instituto de F{\'i}sica de Facultad de Ciencias, Igu{\'a} 4225, Montevideo 11400, Uruguay.}
\author{Nicol{\'a}s Rubido} \email{nrubido@fisica.edu.uy}\email{nicolas.rubidoobrer@abdn.ac.uk}
\affiliation{Universidad de la Rep{\'u}blica, Instituto de F{\'i}sica de Facultad de Ciencias, Igu{\'a} 4225, Montevideo 11400, Uruguay.}
\affiliation{University of Aberdeen, Aberdeen Biomedical Imaging Centre, AB25 2ZG Aberdeen, United Kingdom.}
\affiliation{University of Aberdeen, Institute for Complex Systems and Mathematical Biology, AB24 3UE Aberdeen, United Kingdom.}


\date{\today}
\begin{abstract}
Initially, the logistic map became popular as a simplified model for population growth. In spite of its apparent simplicity, as the population growth-rate is increased the map exhibits a broad range of dynamics, which include bifurcation cascades going from periodic to chaotic solutions. Studying coupled maps allows to identify other qualitative changes in the collective dynamics, such as pattern formations or hysteresis. Particularly, hysteresis is the appearance of different attracting sets, a set when the control parameter is increased and another set when it is decreased -- a multi-stable region. In this work, we present an experimental study on the bifurcations and hysteresis of nearly identical, coupled, logistic maps. Our logistic maps are an electronic system that has a discrete-time evolution with a high signal-to-noise ratio ($\sim10^6$), resulting in simple, precise, and reliable experimental manipulations, which include the design of a modifiable diffusive coupling configuration circuit. We find that the characterisations of the isolated and coupled logistic-maps' dynamics agrees excellently with the theoretical and numerical predictions (such as the critical bifurcation points and Feigenbaum's bifurcation velocity). Here, we report multi-stable regions appearing robustly across configurations, even though our configurations had parameter mismatch (which we measure directly from the components of the circuit and also infer from the resultant dynamics for each map) and were unavoidably affected by electronic noise.
\end{abstract} 



\maketitle
%
\section{Introduction}
The logistic map is a paradigmatic dynamical system that was proposed by May \cite{May} as a model for population growth. It became popular because of its simplicity -- being one-dimensional, having a single parameter, and using a smooth quadratic function --, but also because of its broad dynamical regimes -- going from periodic orbits to chaos. As the growth rate parameter is increased, the map dynamics shows bifurcation cascades with a fractal structure, which are shown to be universal \cite{Feigenbaum,Grassberger,Omelchenko,Dodds}. From its initial framework as a population model, the logistic map has been studied thoroughly and used vastly. For example, as an ecological model \cite{Stone, Storch} an encryption machine \cite{Kocarev,Pareek,Mazloom,Singh}, and a noise generator \cite{Phatak,McGonigal}, to name a few. Furthermore, in order to explain the emergence of collective phenomena from a tractable framework, coupled logistic maps have been studied numerically \cite{Kaneko,Wang,Vandermeer} to explain chaotic synchronisation \cite{Viana,Marti,CMasoller,Xie} or model effects of diversity and heterogeneity in competing populations \cite{Lloyd}, to name a few.

The study of coupled maps allows to identify other qualitative changes in the collective dynamics, such as hysteresis \cite{Neufeld,Boccaletti,Gu}, or pattern formations, such as chimeras \cite{Uenohara,Morie,Meena}. Particularly, hysteresis corresponds to having a different attracting set when the control parameter is increased than when it is decreased (a phenomenon that is fairly known in ferromagnetic materials). For example, coupled map lattices with a one-humped chaotic map and an unstable Laplacian coupling show hysteresis when observed as the control parameter is changed \cite{Neufeld}. Similarly, coupled logistic maps under a fixed multiplicative coupling also show hysteresis when the logistic parameter is tuned \cite{Gu}. However, these and other works solely study hysteresis -- and other kinds of crisis -- by means of numerical simulations, without taking into account parameter mismatch or the role of intrinsic noise. Moreover, to the best of our knowledge, we are still unaware on how persistent the hysteresis is when changes in the coupling configuration are introduced. Namely, what is the dependence between having hysterical behaviour and the particular coupling configuration chosen.

In this work, we implement a low-cost electronic circuit that models diffusely-coupled logistic maps and report its emerging bifurcations and hysteresis as the coupling strength is changed. Our experimental set-up is based on the logistic map circuit we define in Ref.~\cite{LHer}, which we now extend to include a circuit board that allows to change the coupling configuration (i.e., the connectivity between maps) as well as the number of interacting maps. The circuit allows precise and reliable manipulations (with an average $1\%$ parameter uncertainty per map) with high signal-to-noise ratio ($\sim 10^6$). Our results are centred on $6$ logistic maps, where we report hysteresis as a function of coupling strength for $52$ different configurations, i.e., $52$ networks. We show that hysteresis appears robustly across configurations but for different coupling strength regions -- in spite of small parameter mismatch and electronic noise.
%
\section{Model and Methods}
Our results are obtained from experimentally implementing $6$ Kaneko-coupled \cite{Kaneko} logistic maps. The equations of motion for the $i$-th map is given by
\begin{equation}
x_{n+1}^{(i)} = (1 - \varepsilon) f(x_n^{(i)};r_i) + \varepsilon \sum_{j=1}^{N} \dfrac{A_{ij}}{d_i} f(x_n^{(j)};r_j),
\label{eq_KanekoLogMaps}
\end{equation}
where $x_{n}^{(i)}$ [$x_{n+1}^{(i)}$] is the $i$-th map state ($i = 1,\ldots,6$) at iteration $n$ [$n+1$] (with $n\geq0$, $x_0^{(i)}$ being the initial condition), $f(x;r) \equiv r x (1 - x)$ is the logistic function with parameter $r$, and $\varepsilon$ is the coupling strength, which acts as our global control parameter. The adjacency matrix, $\mathbb{A}$, defines the coupling configurations, which we assume bidirectional and unweighted, and define the number of neighbours (node degree) that each map has in the configuration: $d_i = \sum_{j} A_{ij}$ for $i = 1,\ldots,6$. Its binary entries, $A_{ij}$, indicated whether map $i$ and $j$ are connected, $A_{ij} = 1$, or disconnected, $A_{ij} = 0$.

\begin{figure}[htbp]
 \begin{center}
    \includegraphics[width=0.9\columnwidth]{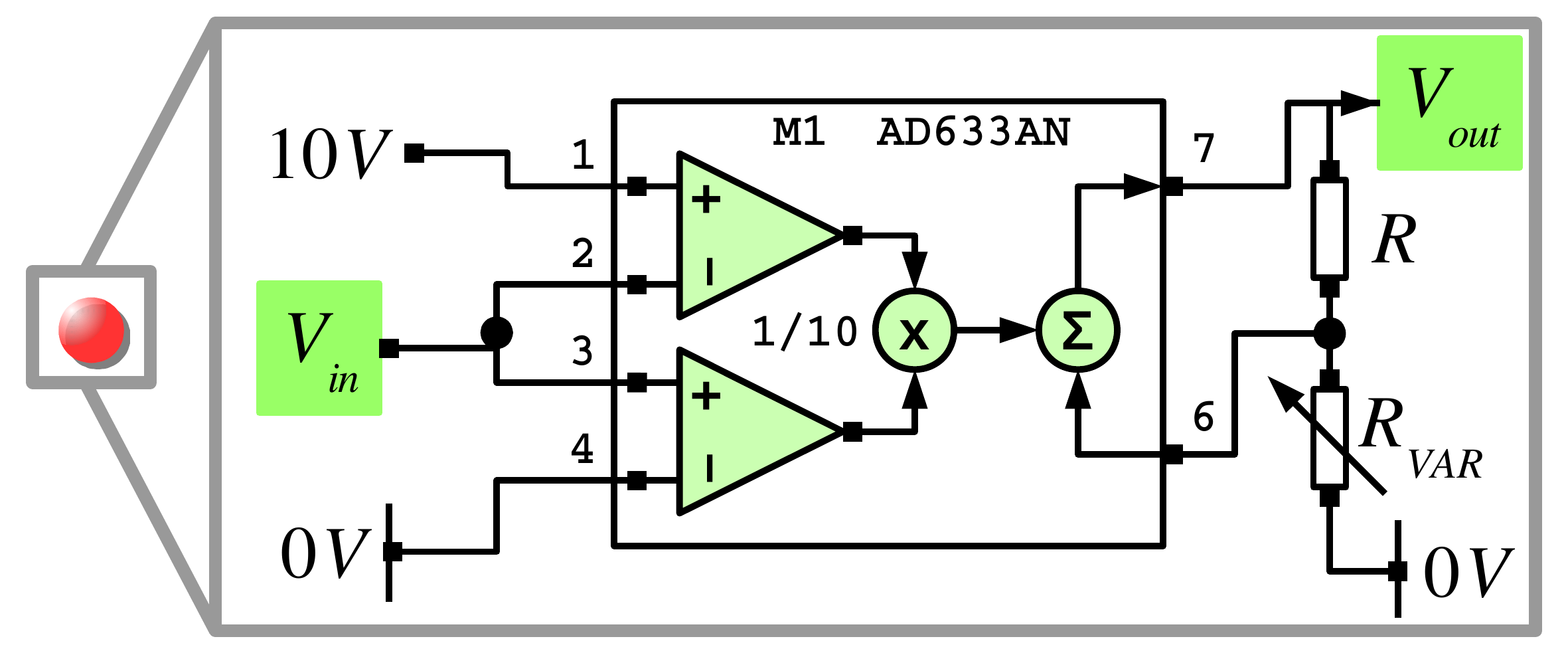}\\
    \includegraphics[width=0.32\columnwidth]{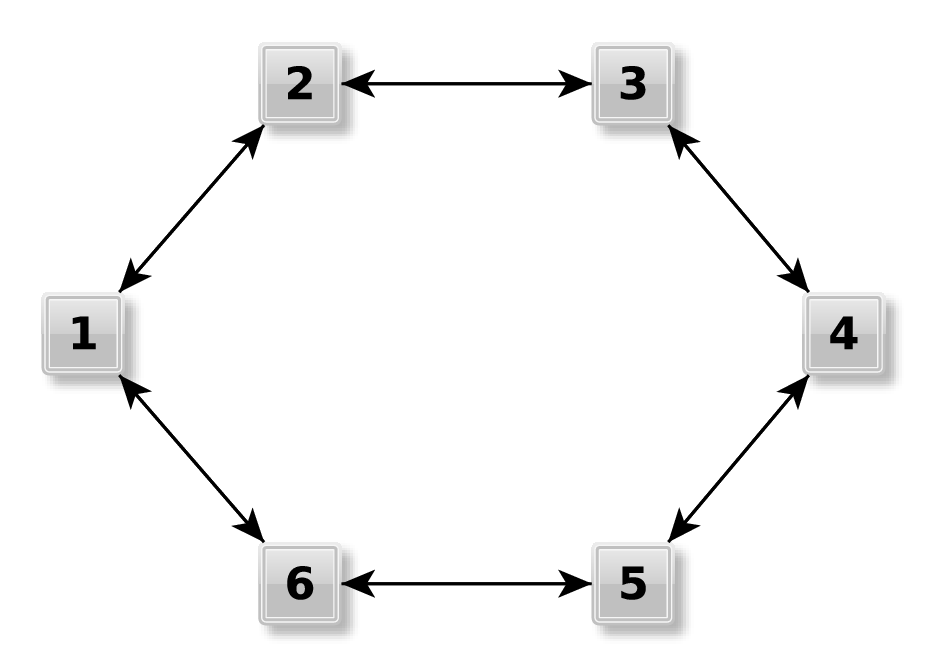}
    \includegraphics[width=0.32\columnwidth]{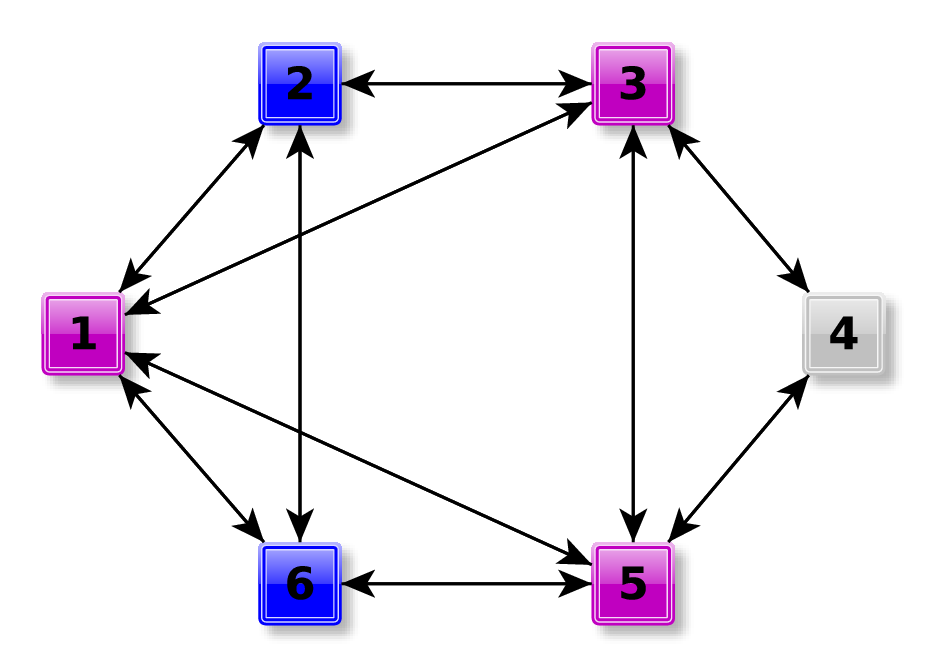}
    \includegraphics[width=0.32\columnwidth]{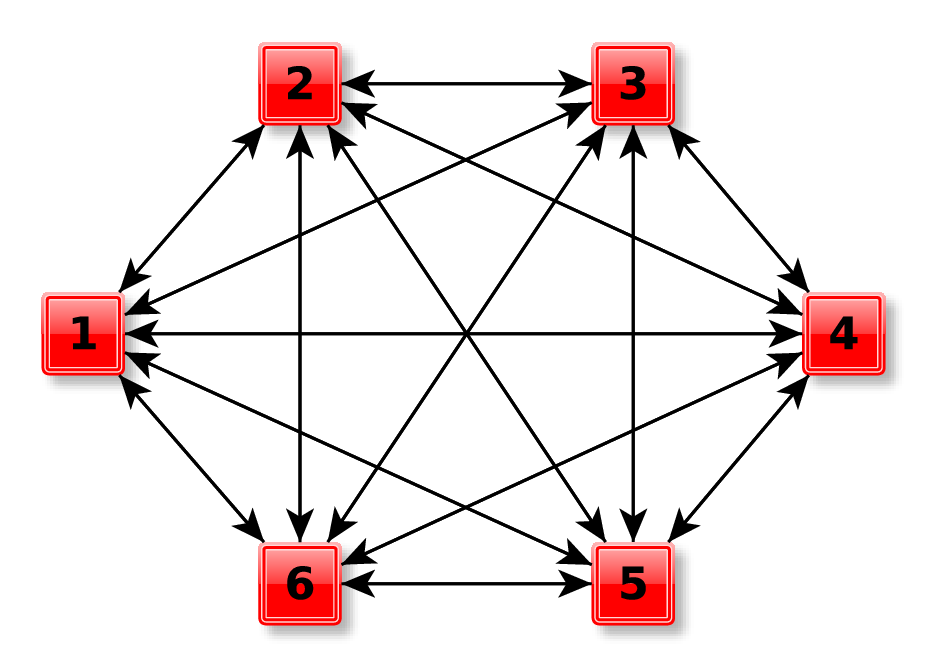}
 \end{center}\vspace{-0.8pc}
 \caption{Schematic representation of the logistic function circuit (top panel) and coupling configurations (bottom panels). The colours for each map in the configurations are used to differentiate their number of connections. Each logistic circuit is attached to a sample-and-hold circuit through $V_{out}$, which creates time-sustained states for the output voltages (i.e., holds the voltage output constant for a fixed time-window). A coupling block then entangles all map's output by a feedback through $V_{in}$; see \ref{App_Blocks}}
    \label{fig_ExpLogMapNets}
\end{figure}

In our experimental setting -- schematically shown in Fig.~\ref{fig_ExpLogMapNets} --, the map parameters, $r_i$, are adjustable by resistors, $r_{exp} = 1 + R_{var}/R$, up to a $1\%$ precision (according to the manufacturer), where $R = 1\,k\Omega$. The logistic function, $f(x;r)$, is implemented by an analog multiplier AD633AN (top panel in Fig.~\ref{fig_ExpLogMapNets}) up to an error less than $3\%$ with respect to the theoretical quadratic function. Its discrete-time evolution is implemented by a sample-and-hold circuit \cite{LHer}, which holds the voltage output, $V_{out}$, constant for a fixed time-window, before releasing it to our coupling block circuit; details in \ref{App_Blocks} The resultant configurations are adaptable, as shown in the bottom panels of Fig.~\ref{fig_ExpLogMapNets}. We control $\varepsilon = V_{NI}/10\,V$, by the analog output of a National Instrument Data Acquisition (NIDAQ) USB 6216, which has $3.5\,\mu V$ precision, making $\varepsilon$'s uncertainty lower than $10^{-6}$. In particular, we choose to change $\varepsilon$ in increments (or decrements) of $\Delta\varepsilon = 1/256$.

Aside being able to tune the logistic function parameter from the resistors that define the circuit, $r_{exp}$, we also infer their effective values from the recorded signals of the isolated maps. We note these parameters as $r_i$, which effectively constitute the parameters of the circuit's mapping. The reason behind this choice is that, even when tuning the resistors to be identical (within the precision set by the manufacturer), the remaining circuit components and inherent uncertainties make each map to behave slightly different. Specifically, the isolated logistic maps (i.e., when $\varepsilon = 0$) are tuned to be in the chaotic regime, $r_{exp} = 3.80\pm0.03$, and respond according to the following map parameters: $r_1 = 3.7364$, $r_2 = 3.7537$, $r_3 = 3.7609$, $r_4 = 3.7446$, $r_5 = 3.7298$, and $r_6 = 3.7300$, with a common uncertainty of $3\times10^{-4}$. We obtain these parameters by making a robust regression between the observed signal, $\{x_n\}_{n = 0}^T$, and the logistic function, $\{x_n\,(1 - x_n)\}_{n=0}^T$. Moreover, we find that their critical bifurcation points miss by less than $8\%$ Feigenbaum's bifurcation-velocity value, $\delta = 4.6692\ldots$ \cite{Feigenbaum}, meaning that the whole bifurcation cascade is reliably reproduced.
 
In what follows, we register signals for $6$ coupled logistic maps, accounting $T = 2\times10^4$ iterations (after discarding $\gtrsim 10^3$ iterations as transient dynamics) and $52$ coupling configurations. Specifically, we explored $52$ adjacency matrices, i.e., $A_{ij}$ in Eq.~\eqref{eq_KanekoLogMaps}. These $A_{ij}$ are constructed from a complete network ($A_{ij} = 1 - \delta_{i,j}\;\forall\,i,j$) by continuously removing links up to the ring network ($A_{ij} = \delta_{i,j=i\pm1}$). This is done disregarding symmetrical configurations, namely, those that can be obtained from relabelling the maps; details in \ref{App_Configs} Our analyses are carried on the resultant stationary discrete-time signals, $\{x_n^{(i)}\}_{n=1}^T$, where the map parameters, $\{r_i\}_{i = 1}^N$, are fixed and the coupling strength, $\varepsilon$, is changed by increments (or decrements) of $\Delta\varepsilon = 1/256$ from $\varepsilon = 0$ (chaotic isolated dynamics) up to $1$. Given our circuitry, the initial conditions are uncontrolled, but keep some memory of the previous stationary state. This is important for the construction of bifurcation diagrams with hysterical behaviour. For example, when we increase [decrease] the coupling, $\varepsilon + \Delta\varepsilon$ [$\varepsilon - \Delta\varepsilon$], we want the system to be close to the previous attractor at $\varepsilon$. This a methodology that is commonly used to construct numerical bifurcation diagrams.
%
\section{Results}
\begin{figure}[htbp]
    \centering
    \includegraphics[width=0.95\columnwidth]{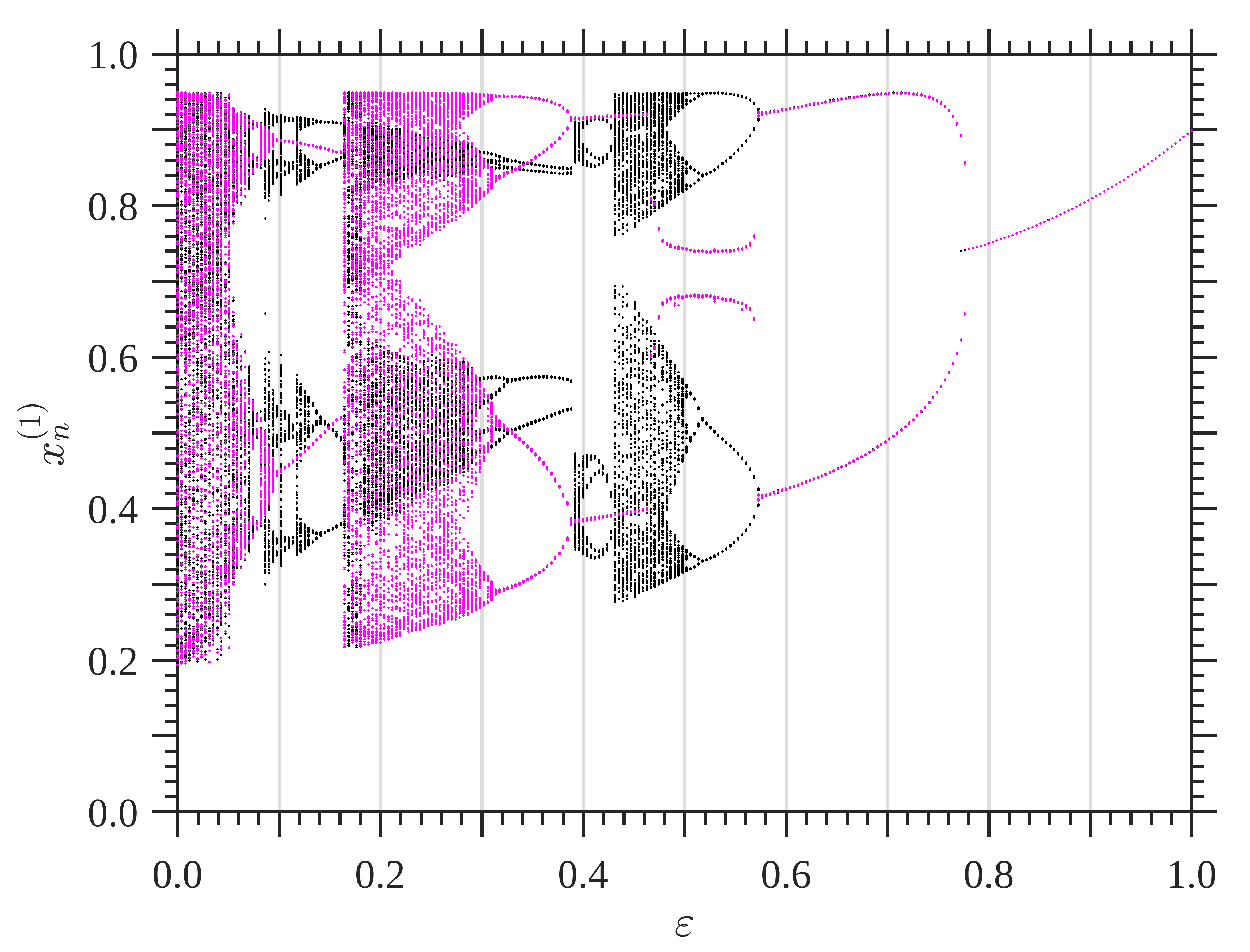} \vspace{-0.6pc}
    \caption{Bifurcation diagram for $1$ logistic map from an experimental system of $6$ nearly-identical maps coupled in a ring (nearest-neighbours). This panel shows the projection of the $6$-dimensional bifurcation diagram. The map parameters are $r_1 = 3.7364$, $r_2 = 3.7537$, $r_3 = 3.7609$, $r_4 = 3.7446$, $r_5 = 3.7298$, and $r_6 = 3.7300$ (with a $3\times10^{-4}$ uncertainty). The coupling strength, $\varepsilon$, is increased (magenta) from $0$ to $1$ and then decreased (black) from $1$ to $0$.}
    \label{fig_BifExample}
\end{figure}

We study the bifurcations of $6$ experimentally coupled logistic maps for $52$ coupling configurations (i.e., networks) as a function of the coupling strength, $\varepsilon$, when it is increased or decreased. We note that the resultant bifurcation cascades follow the known period-doubling route for all configurations, but in an inverse way. Generally, for most configurations, as $\varepsilon$ is increased from $0$ to $1$, the system exhibits an aperiodic regime (chaotic isolated dynamics) until $\varepsilon \lesssim 0.05$, then successively decreasing its periodic behaviour -- going through nearly-synchronous regions -- up to a fixed point appearing after $\varepsilon \gtrsim 0.8$. We can see this from Fig.~\ref{fig_BifExample}, where we show the resultant bifurcation diagram for map $1$ in a ring configuration (bottom left panel in Fig.~\ref{fig_ExpLogMapNets}).

More importantly, our analyses show that hysteresis appears in all configurations explored, although it appears at different $\varepsilon$ depending on the configuration and with different branching characteristics -- showing different multi-stable states. For example, in Fig.~\ref{fig_BifExample}, we can see that there is a hysterical region at $0.1 \lesssim \varepsilon \lesssim 0.55$, where the increasing (magenta) and decreasing (black) of $\varepsilon$ results in different dynamics. As expected, hysteresis reveals multi-stable regions of the coupled logistic maps. Here, we are showing that it appears robustly across configurations and emerges in spite of the experimentally inherent parameter mismatch and electronic noise present in the system.

\begin{figure}[htbp]
    \centering
    \includegraphics[width=0.95\columnwidth]{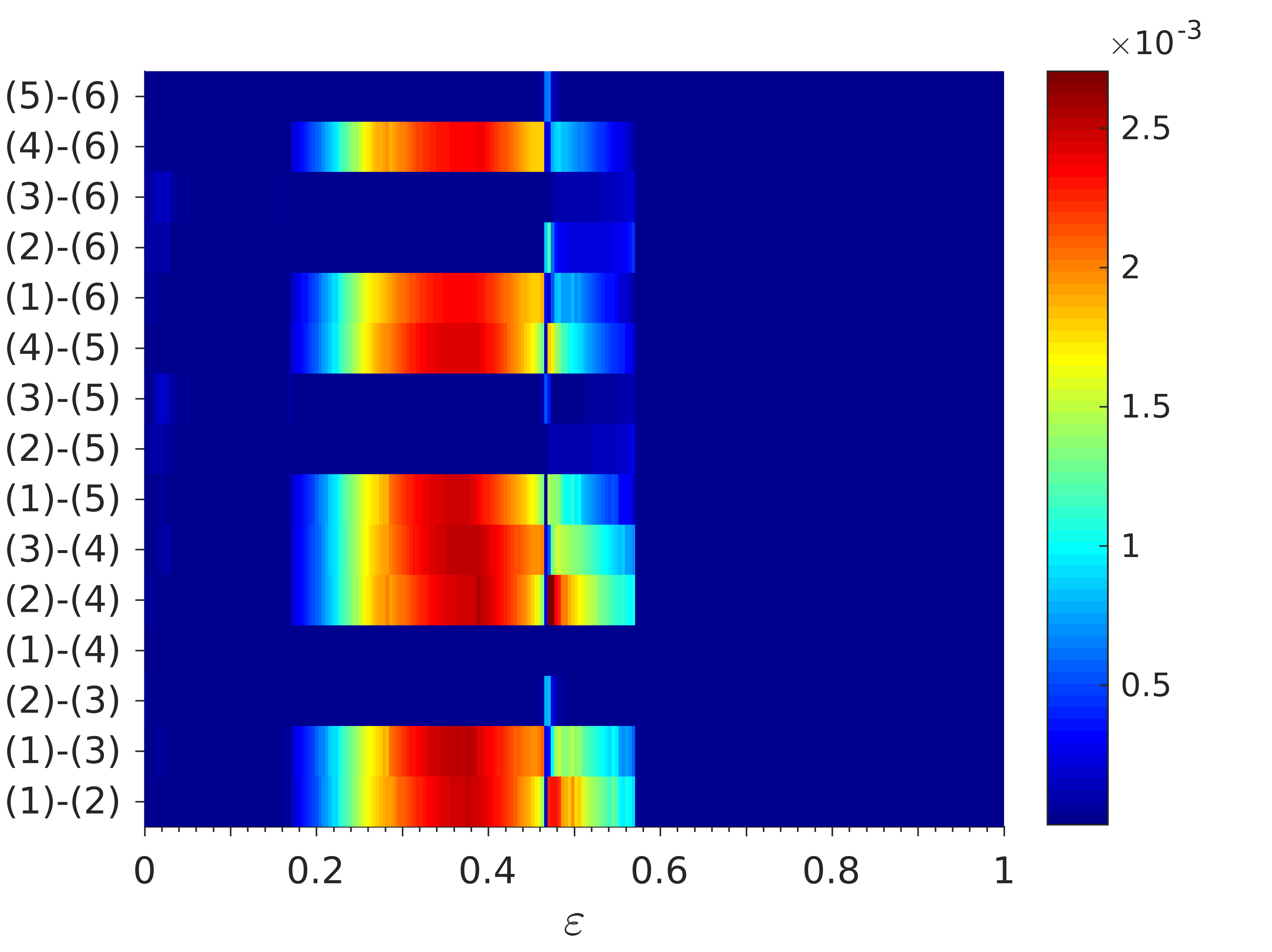} \vspace{-0.6pc}
    \caption{Pair-wise variance, $\sigma_{ij}^2$, for the coupled logistic maps in Fig.~\ref{fig_BifExample} as the coupling strength, $\varepsilon$, is increased. The colours are these time-averaged quadratic differences for all $i$ and $j$ (map) signals.}
    \label{fig_VarExample}
\end{figure}

We group the hysteresis results according to the similarities that the order parameter exhibits for the different configurations, which we define as the average pair-wise variances. For example, Fig.~\ref{fig_VarExample} shows $\sigma^2_{ij}$ for the ring configuration of Fig.~\ref{fig_BifExample} when we increase $\varepsilon$ (magenta). In what follows, our order parameter, $\overline{\sigma^2}$, is the average of these $\sigma^2_{ij}$, considering all ordered $(i,j)$ pairs; see details in \ref{App_Param} In general, we note that the largest [smallest] hysterical loop is found for the ring [complete] network, i.e., when we connect $2$ neighbouring [all-to-all] maps. In between these two extremes, there are slight variations depending on the coupling configuration, but hysteresis appears in all configurations (approximately) for $0.1\lesssim \varepsilon \lesssim 0.3$.

%
	\subsection{Multi-stable behaviour as the ring configuration}
\begin{figure}[htbp]
    \centering
    \includegraphics[width=0.32\columnwidth]{1.png}
    \includegraphics[width=0.32\columnwidth]{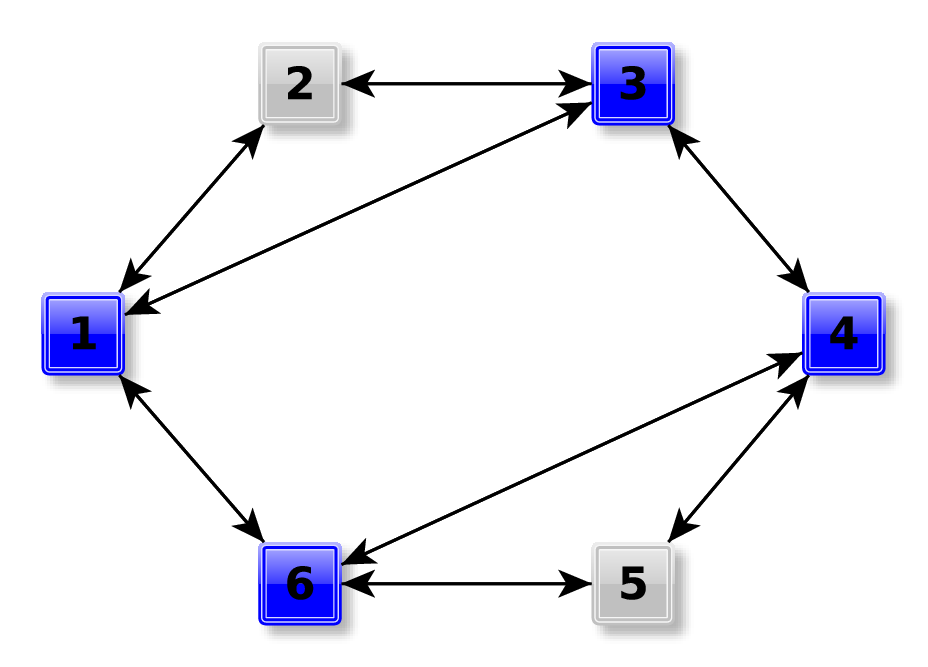}
    \includegraphics[width=0.95\columnwidth]{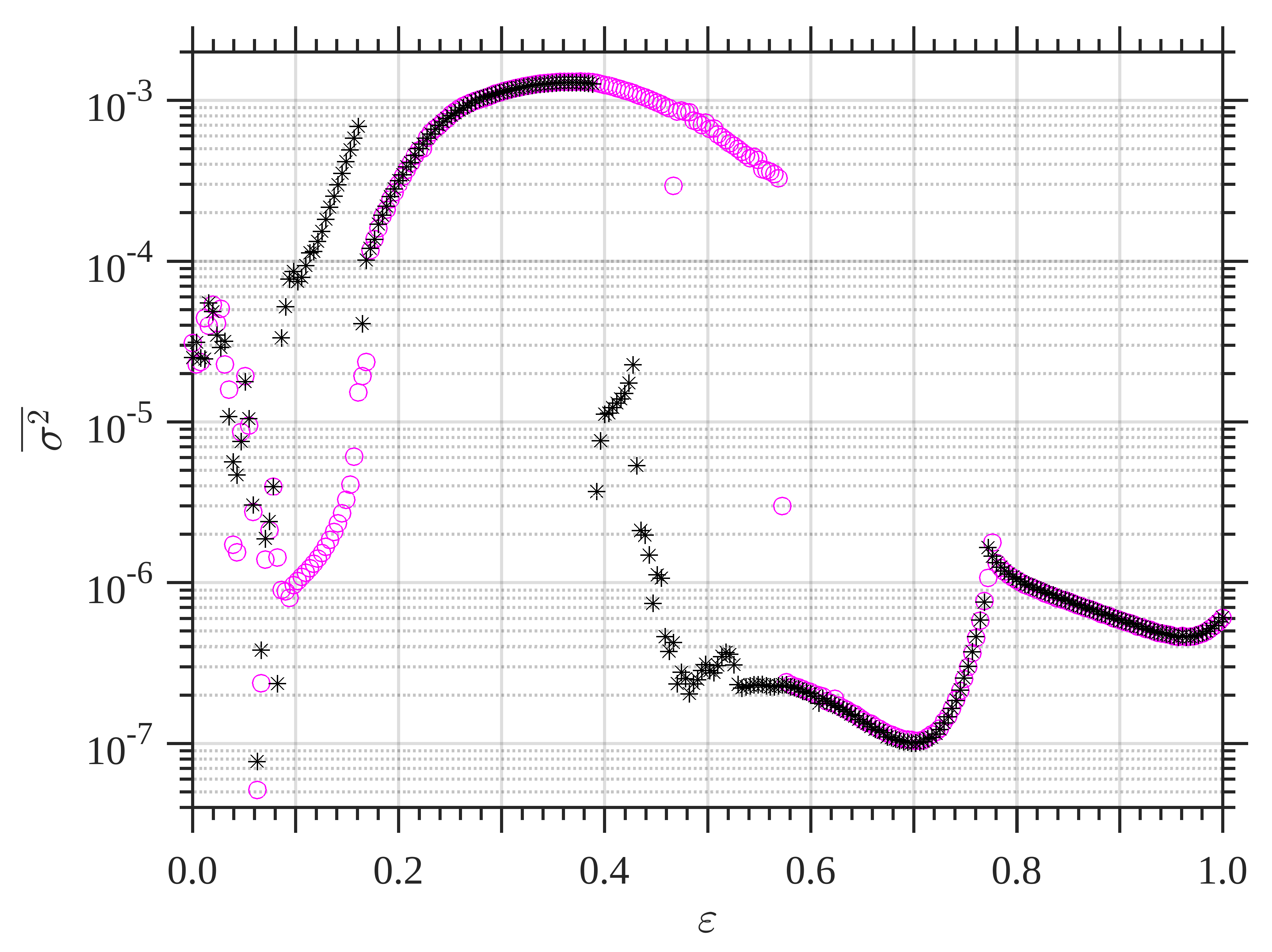} \vspace{-0.5pc}
    \caption{Hysteresis (bottom panel) in ring (top left panel) and symmetrically-perturbed ring (top right panel) configurations of $6$ nearly-identical, electronically-coupled, logistic maps. The coupling strength, $\varepsilon$, is increased (magenta circles) from $0$ to $1$, and subsequently decreased (black asterisks) from $1$ to $0$. Multi-stability is revealed by the forward and backward differences in the order parameter, $\overline{\sigma^2}$, which is the pair-wise averaged variance of the $6$ coupled logistic maps from Figs.~\ref{fig_BifExample} and \ref{fig_VarExample}.}
    \label{fig_RingNets}
\end{figure}

The largest hysteresis we find using $\overline{\sigma^2}$ is for the ring configuration (top left panel in Fig.~\ref{fig_RingNets}). The different collective dynamics that this system has when increasing (magenta circles) or decreasing (black asterisks) $\varepsilon$ span a multi-stable region, which is found at $0.1\lesssim\varepsilon\lesssim 0.6$ and has chaotic attractors and periodic windows emerging and collapsing after period-doubling cascades. For this ring configuration, we show in the bottom panel of Fig.~\ref{fig_RingNets} the order parameter, $\overline{\sigma^2}$ as a function of the coupling strength, $\varepsilon$. In general, we highlight that whenever $\overline{\sigma^2} \simeq 0$, the coupled maps are nearly completely synchronous, i.e., $x_i(t) \simeq x_j(t)$ $\forall\,t$ and $\forall\,i,j$. In other words, $\overline{\sigma^2}$ reveals the average distance of the $6$ dimensional (coupled maps) trajectory to the completely synchronous manifold, i.e., $x_i(t) = x_j(t)$.

From analysing the $52$ configurations, we note that the hysterical behaviour of the ring configuration is approximately similar to only $1$ other configuration, namely, the symmetrically-perturbed configuration that we show in the top right panel of Fig.~\ref{fig_RingNets}. Both configurations result in a  $\overline{\sigma^2}$ behaviour similar to that of Fig.~\ref{fig_RingNets} bottom panel, where increasing (magenta circles) and decreasing (black asterisks) $\varepsilon$ shows a hysteresis for $0.08 \lesssim\varepsilon\lesssim 0.17$ and $0.4 \lesssim\varepsilon\lesssim 0.6$. In fact, we can see from the bifurcation diagram of Fig.~\ref{fig_BifExample} that there is also a multi-stable region between these $\varepsilon$ regions at $0.17 \lesssim\varepsilon\lesssim 0.4$. However, we miss this in-between region when using $\overline{\sigma^2}$. These results show that the $2$ configurations have $3$ multi-stable regions: $2$ regions where the system goes through different attracting sets, which are closer to the completely-synchronous manifold either when increasing or decreasing $\varepsilon$ (the $2$ hysterical loops in Fig.~\ref{fig_RingNets}), and $1$ region where the system has different attracting sets with an heterogeneous behaviours per map, but are symmetric with respect to increasing or decreasing $\varepsilon$, namely, the $0.17 \lesssim\varepsilon\lesssim 0.4$ curves in Fig.~\ref{fig_RingNets}.
%
	\subsection{Multi-stable behaviour as the all-to-all configuration}
\begin{figure}[htbp]
    \centering
    \includegraphics[width=0.32\columnwidth]{52.png}
    \includegraphics[width=0.32\columnwidth]{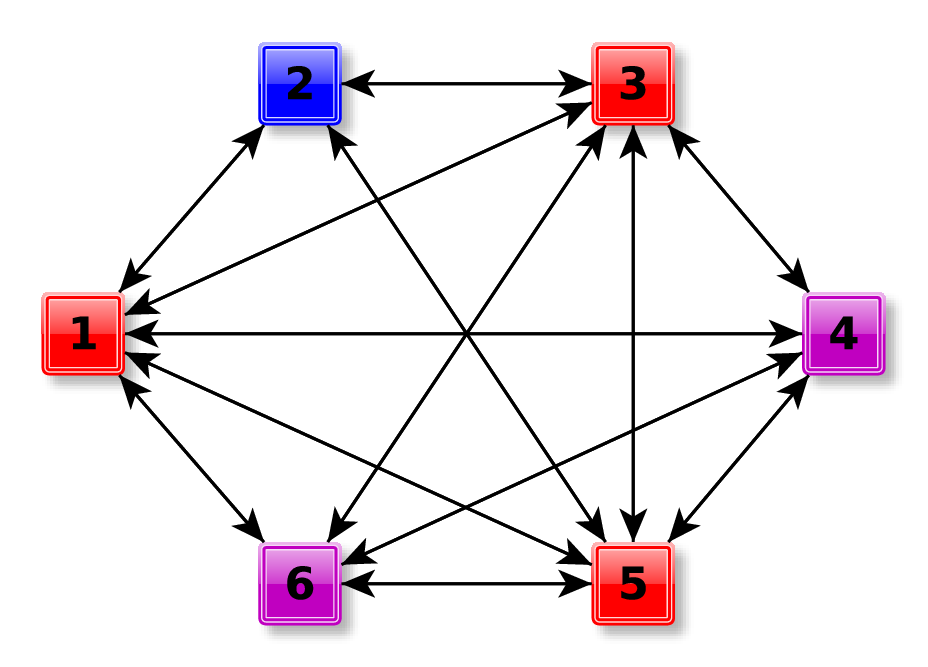}
    \includegraphics[width=0.32\columnwidth]{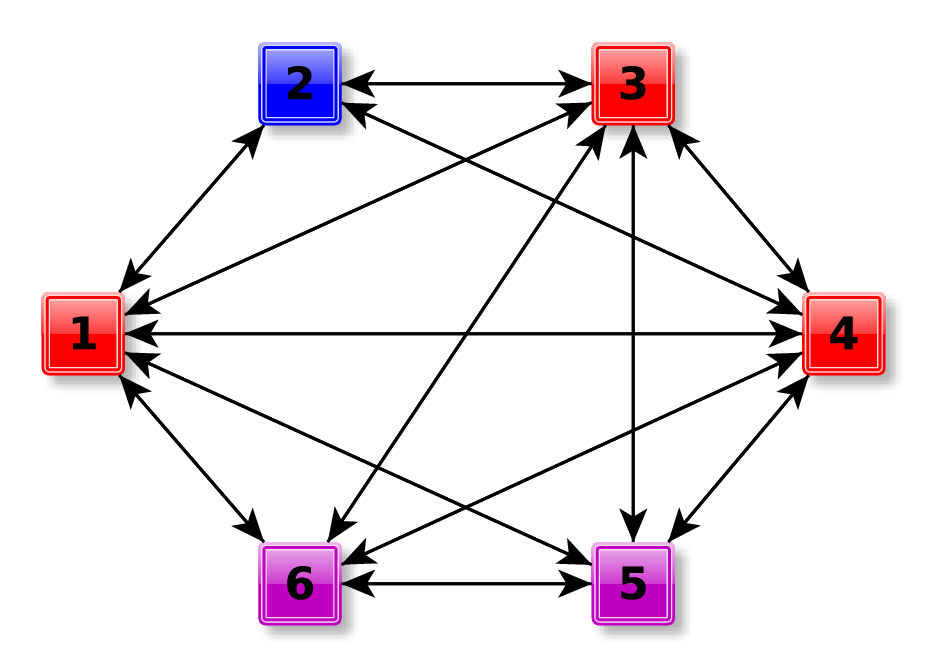}
    \includegraphics[width=0.95\columnwidth]{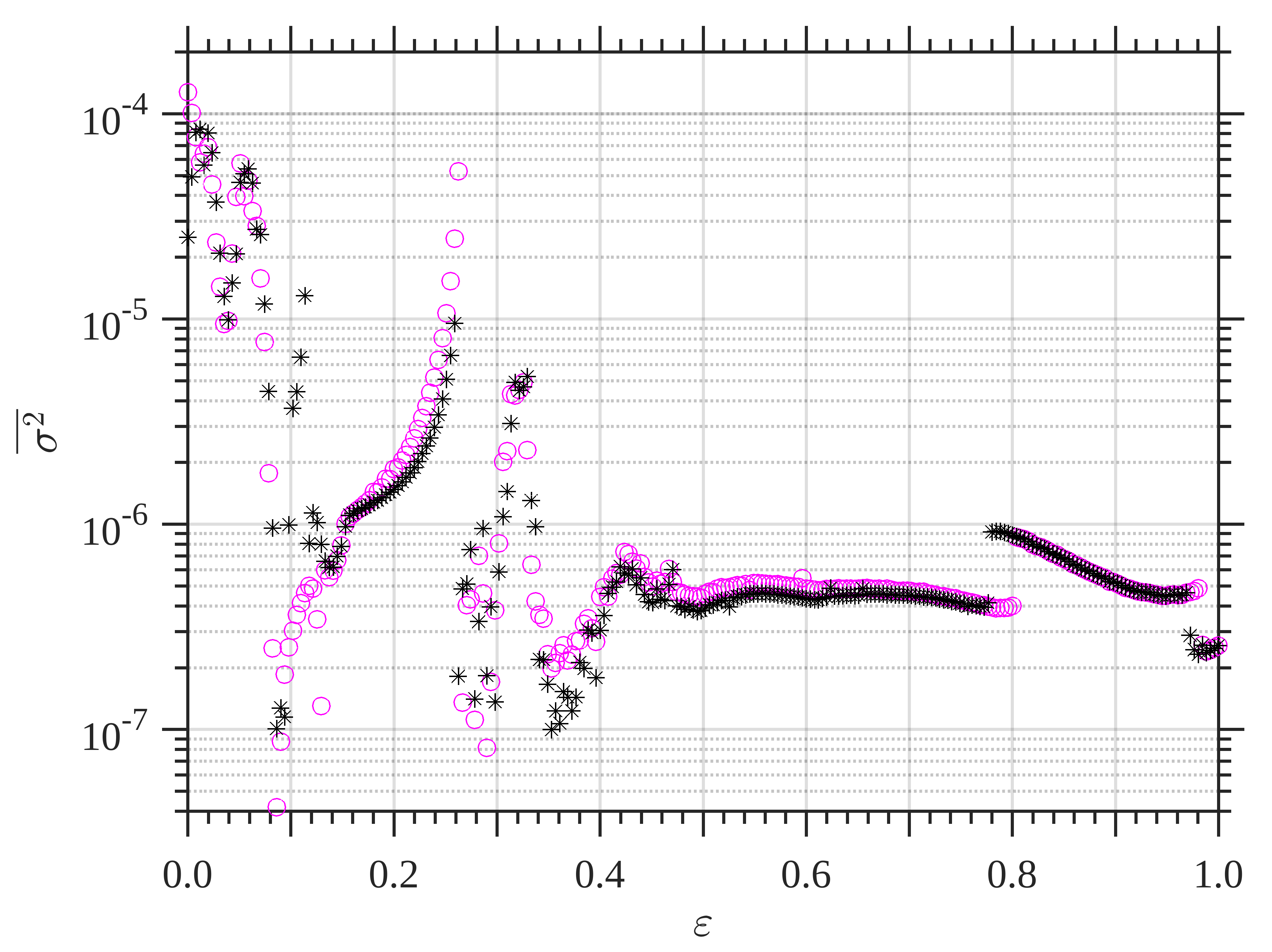} \vspace{-0.5pc}
    \caption{Hysteresis (bottom panel) in all-to-all (top left panel) and perturbed all-to-all (top middle and right panels) configurations of $6$ nearly-identical, experimentally-coupled, logistic maps. Map parameters ($r_i \simeq r = 3.75\pm0.03\,\forall\,i$), control and order parameters ($\varepsilon$ and $\overline{\sigma^2}$, respectively), symbols, and colours are as in Figs.~\ref{fig_BifExample} and \ref{fig_RingNets}.}
    \label{fig_CompleteNets}
\end{figure}

The smallest hysteresis loop we register with $\overline{\sigma^2}$ happens for the all-to-all configuration (top left panel in Fig.~\ref{fig_CompleteNets}), with hysteresis appearing at different $\varepsilon$ (analogously to the ring network). From the bottom panel in Fig.~\ref{fig_CompleteNets}, we can see that $\overline{\sigma^2}$ for this configuration shows a small hysteresis loop around the coupling region of $0.10 \lesssim\varepsilon\lesssim 0.25$; plus, a noisy region for weak coupling-strengths, $0.0 \lesssim\varepsilon\lesssim 0.1$. We highlight that we observe this multi-stable behaviour in $\overline{\sigma^2}$ for $2$ more configurations (out of the remaining $51$), which are shown in the top middle and right panels of Fig.~\ref{fig_CompleteNets}.

In particular, the hysteresis in Fig.~\ref{fig_CompleteNets} appears due to a small-amplitude chaotic attractor and a period $2$ region, which happen at slightly shifted $\varepsilon$, either when it is increased (magenta circles) or decreased (black asterisks). We also note another hysterical region (not noticeable in Fig.~\ref{fig_CompleteNets}) appearing around $\varepsilon\simeq 0.8$, where the system transitions from a period $2$ orbit to a fixed point via a period-doubling bifurcation. Meaning that it takes a larger $\varepsilon$ to collapse to the fixed point when increasing $\varepsilon$ than when decreasing $\varepsilon$. As we discuss in Sect.~\ref{sec_LargeE} for strong couplings, this hysteresis appears robustly across configurations and is close to the complete synchronisation manifold, namely, $\overline{\sigma^2} \lesssim 10^{-6}$.
%
	\subsection{Hysteresis for weak couplings} \label{sec_SmallE}
All the remaining configurations ($47$) show hysteresis for weak couplings; that is, $0.05 \lesssim\varepsilon\lesssim 0.25$. However, as $\varepsilon$ increases, there are branches appearing in the order parameter, $\overline{\sigma^2}$, that differentiate these configurations' collective-dynamics. The branching can be seen for small but increasing $\varepsilon$, where it results in either $1$, $2$, $3$, or $4$ steady growths of $\overline{\sigma^2}$ and subsequent collapses. We note that these branches, which appear within the hysterical region (as well as the hysteresis results in the previous subsections), seem to be uncorrelated with the number of connections per map. In other words, in spite of being able to classify -- for the first time -- these coupling-configurations according to how the system holds a particular $\overline{\sigma^2}$ behaviour, we cannot obtain a general conclusion for the observed branching of $\overline{\sigma^2}$ and the underlying connectivity (or its symmetry). The multi-stable regions appear to be different for different configurations, but are unrelated to the node degrees of each particular network. We show this classification in the following Figs.~\ref{fig_1BranchNets}, \ref{fig_4BranchNets}, \ref{fig_2BranchNets}, and \ref{fig_3BranchNets}, where we group the configurations according to the branching behaviours of $\overline{\sigma^2}$.

As can be seen from the bottom panels in Figs.~\ref{fig_1BranchNets} and \ref{fig_4BranchNets}, a hysteresis loop appears for weak couplings, $0.05 \lesssim\varepsilon\lesssim 0.25$. These $\overline{\sigma^2}$ behaviours correspond to the coupling configuration in the top left panels, however, they are observed (with insignificant changes) for all the configurations shown in the top panels. In the bottom panel of Fig.~\ref{fig_1BranchNets}, we show the representative configuration (corresponding to the top left panel) that exhibits a single upward branch of $\overline{\sigma^2}$ values as $\varepsilon$ is increased, which is nearly reproduced when $\varepsilon$ is decreased. Analogously, but with $4$ upward branches of $\overline{\sigma^2}$ for increasing $\varepsilon$, the bottom panel in Fig.~\ref{fig_4BranchNets} shows the representative configuration that has this branching behaviour for $\overline{\sigma^2}$ (corresponding to the top left panel). We carry this analysis for the other two possible branching behaviours, where Figs.~\ref{fig_2BranchNets} and \ref{fig_3BranchNets} show the resultant $\overline{\sigma^2}$ with $2$ and $3$ branches, respectively, and all the configurations holding these behaviours in $\overline{\sigma^2}$.

\begin{figure}[h!]
    \centering
    \includegraphics[width=0.32\columnwidth]{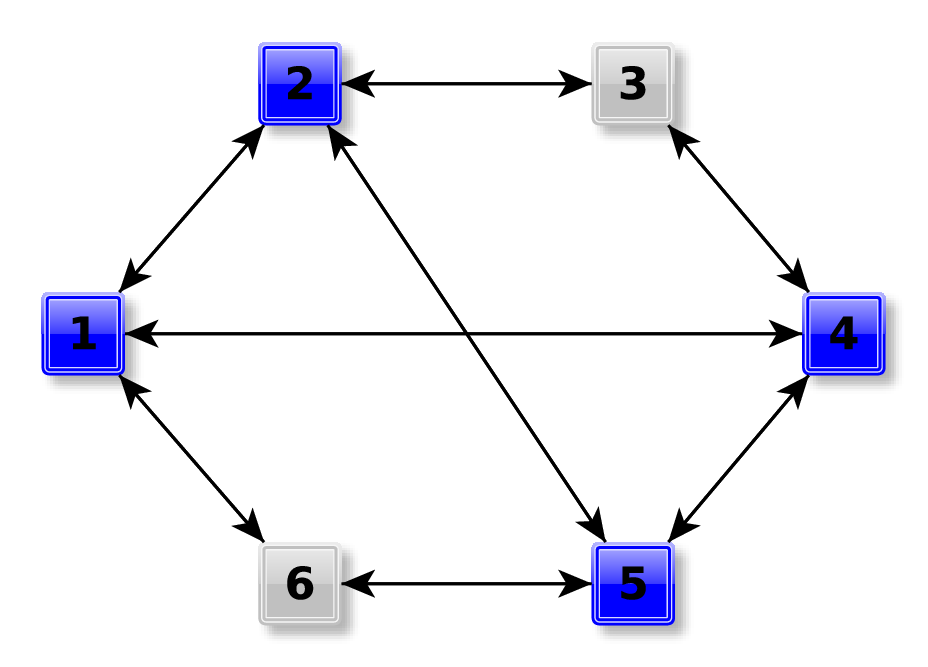}
    \includegraphics[width=0.32\columnwidth]{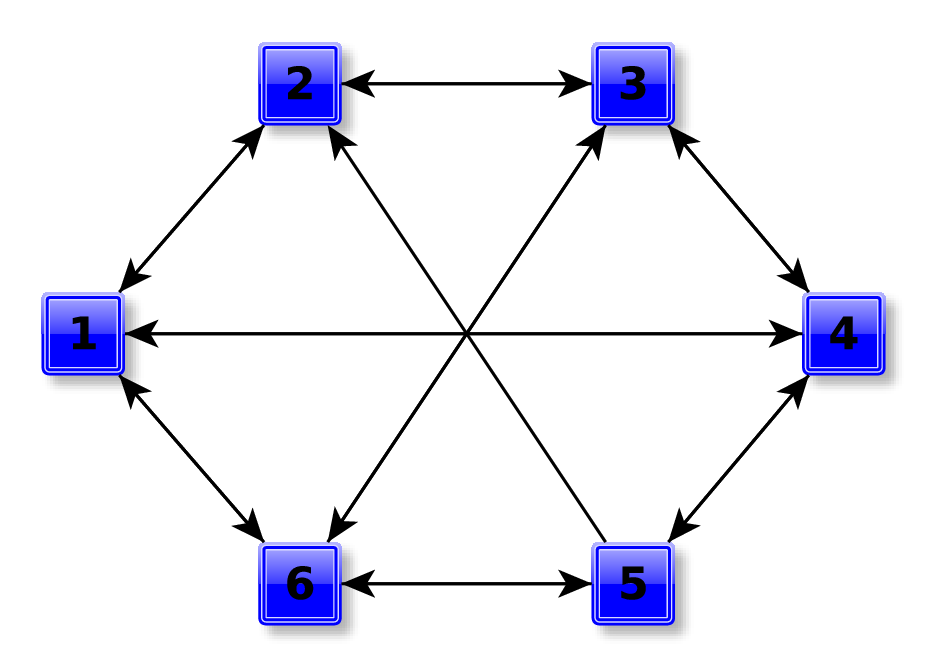}
    \includegraphics[width=0.32\columnwidth]{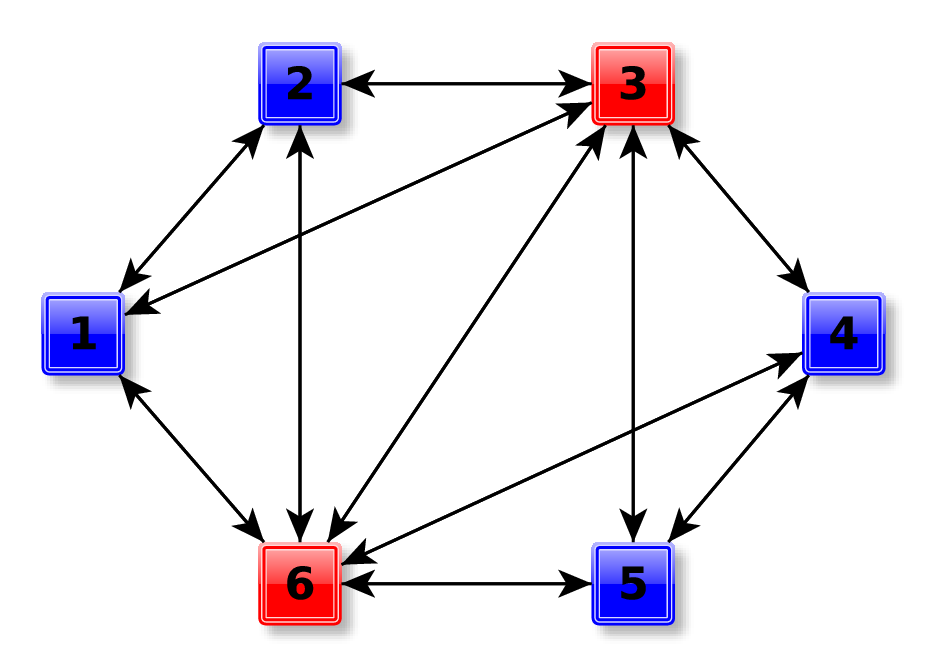}
    \includegraphics[width=0.32\columnwidth]{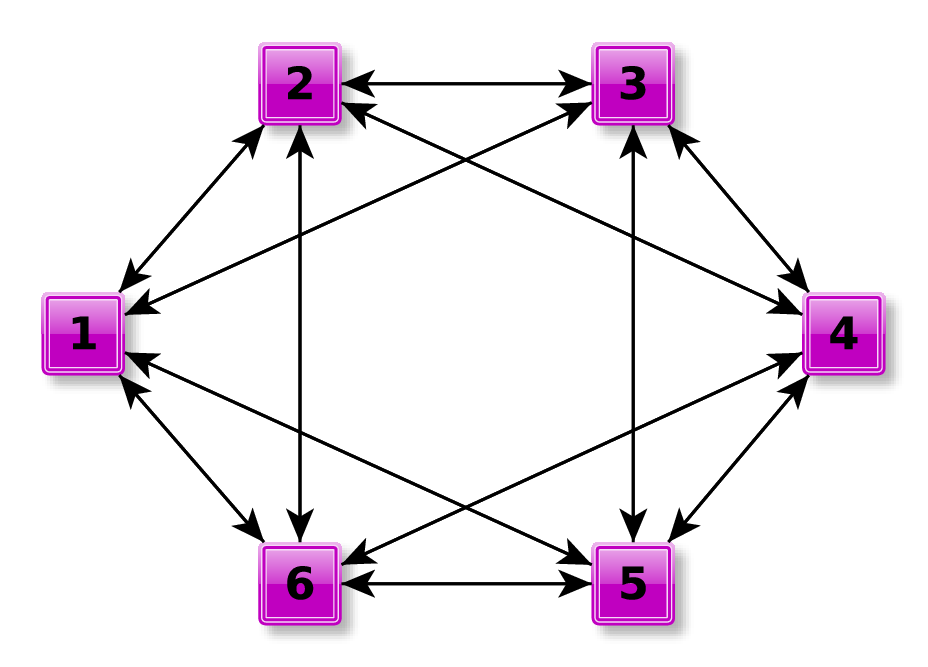}
    \includegraphics[width=0.32\columnwidth]{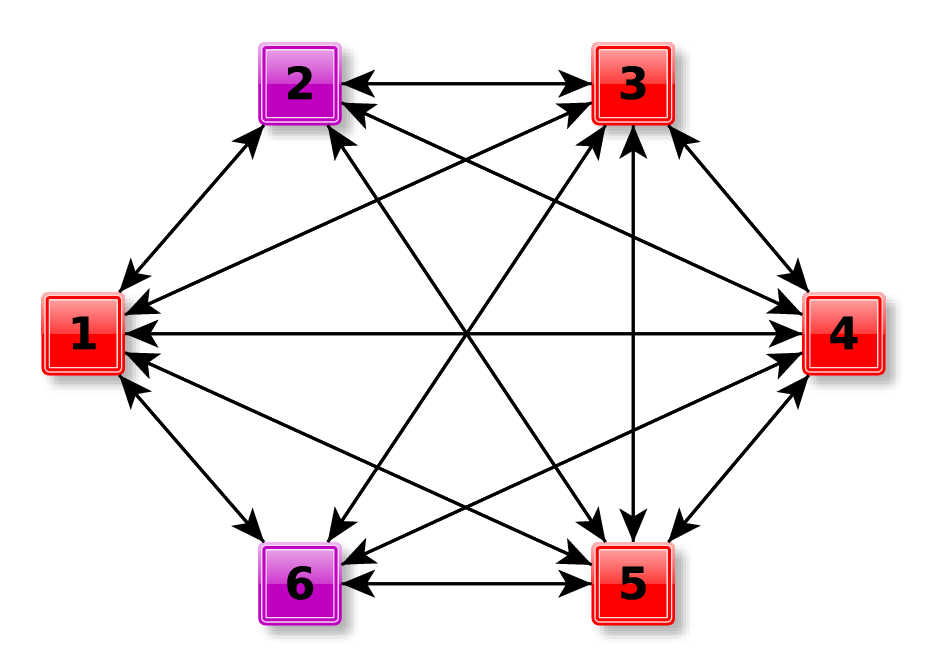}
    \includegraphics[width=0.95\columnwidth]{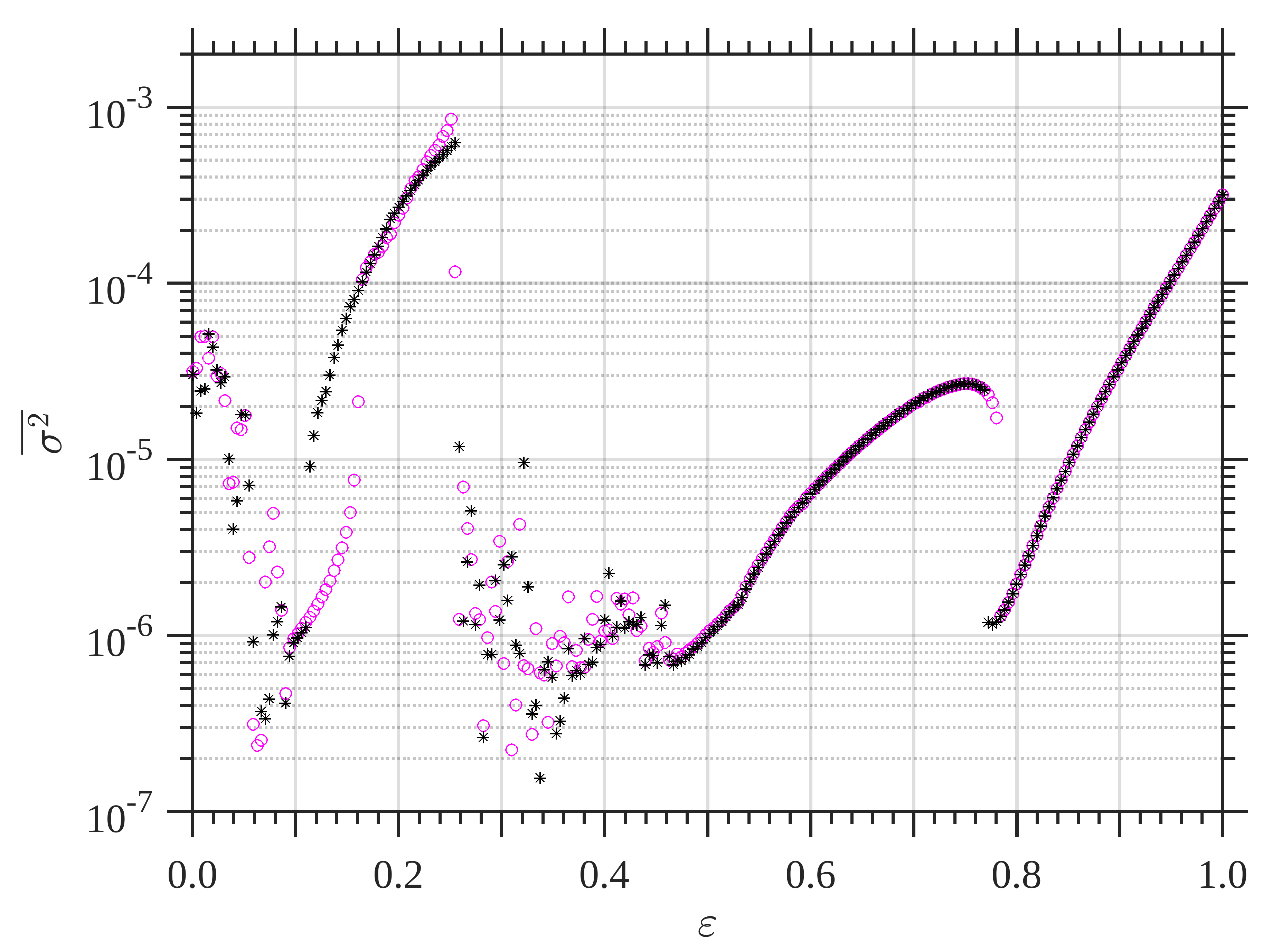} \vspace{-0.5pc}
    \caption{Hysteresis (bottom panel) in different configurations (top panels) of $6$ nearly-identical, coupled, logistic maps. Map parameters ($r_i \simeq r = 3.75\pm0.03\,\forall\,i$), control and order parameters ($\varepsilon$ and $\overline{\sigma^2}$, respectively), symbols, and colours are as in Figs.~\ref{fig_BifExample} and \ref{fig_RingNets}. The $\overline{\sigma^2}$ values correspond to the top left coupling-configuration.}
    \label{fig_1BranchNets}
\end{figure}

\begin{figure}[h!]
    \centering
    \includegraphics[width=0.32\columnwidth]{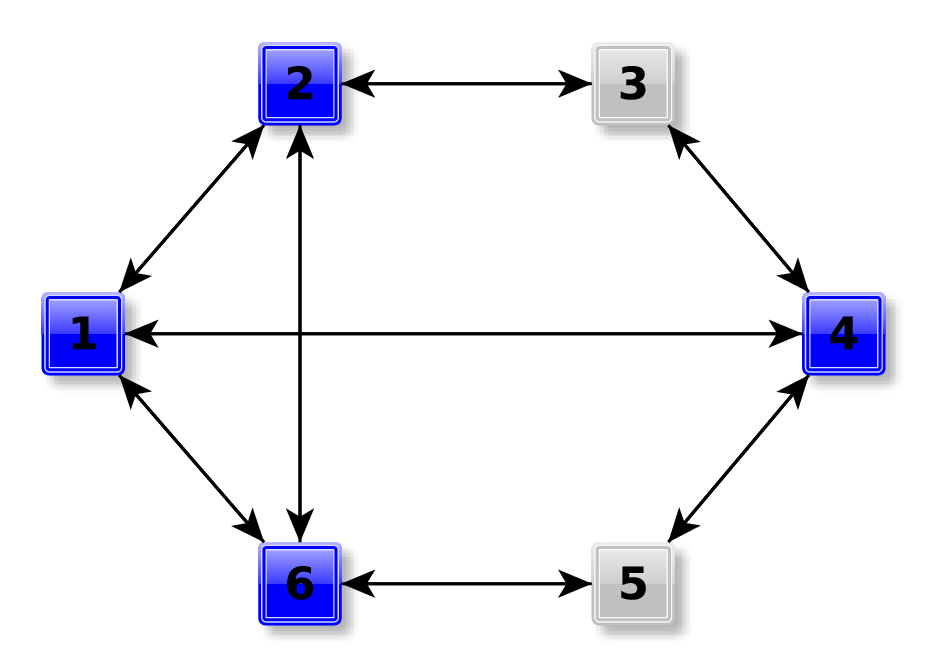}
    \includegraphics[width=0.32\columnwidth]{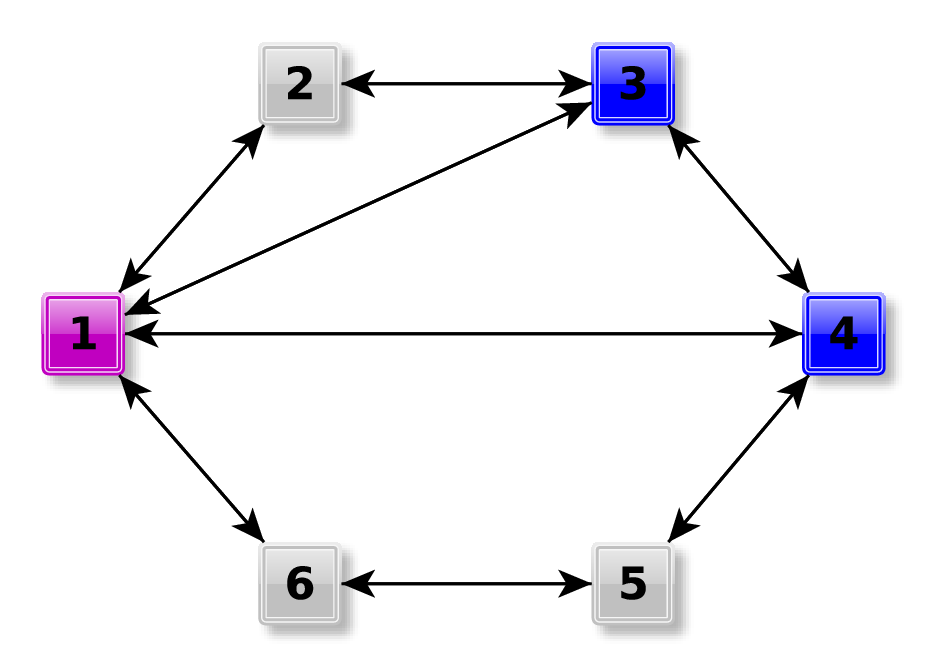}
    \includegraphics[width=0.32\columnwidth]{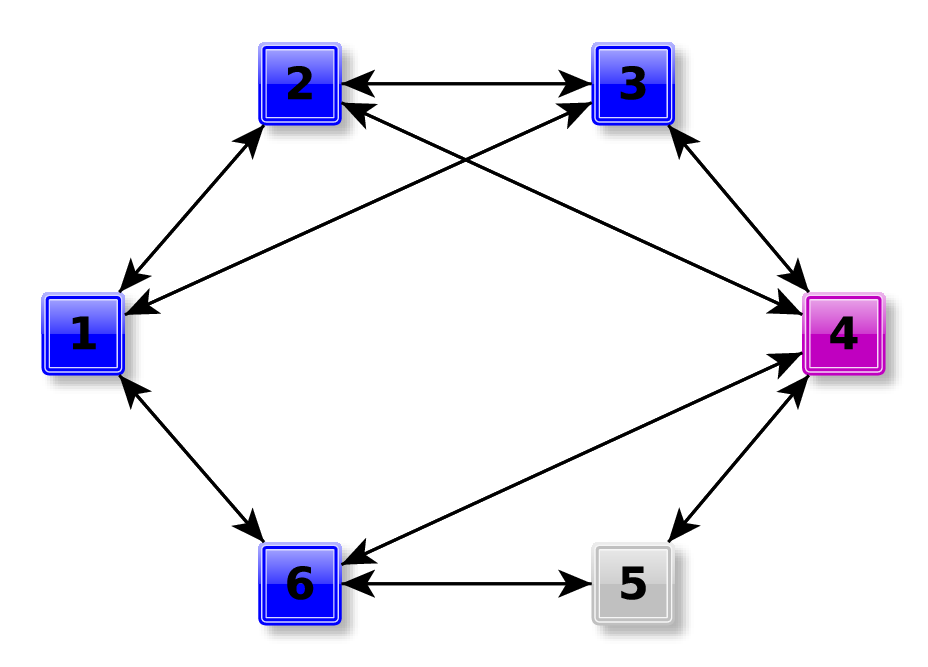}
    \includegraphics[width=0.95\columnwidth]{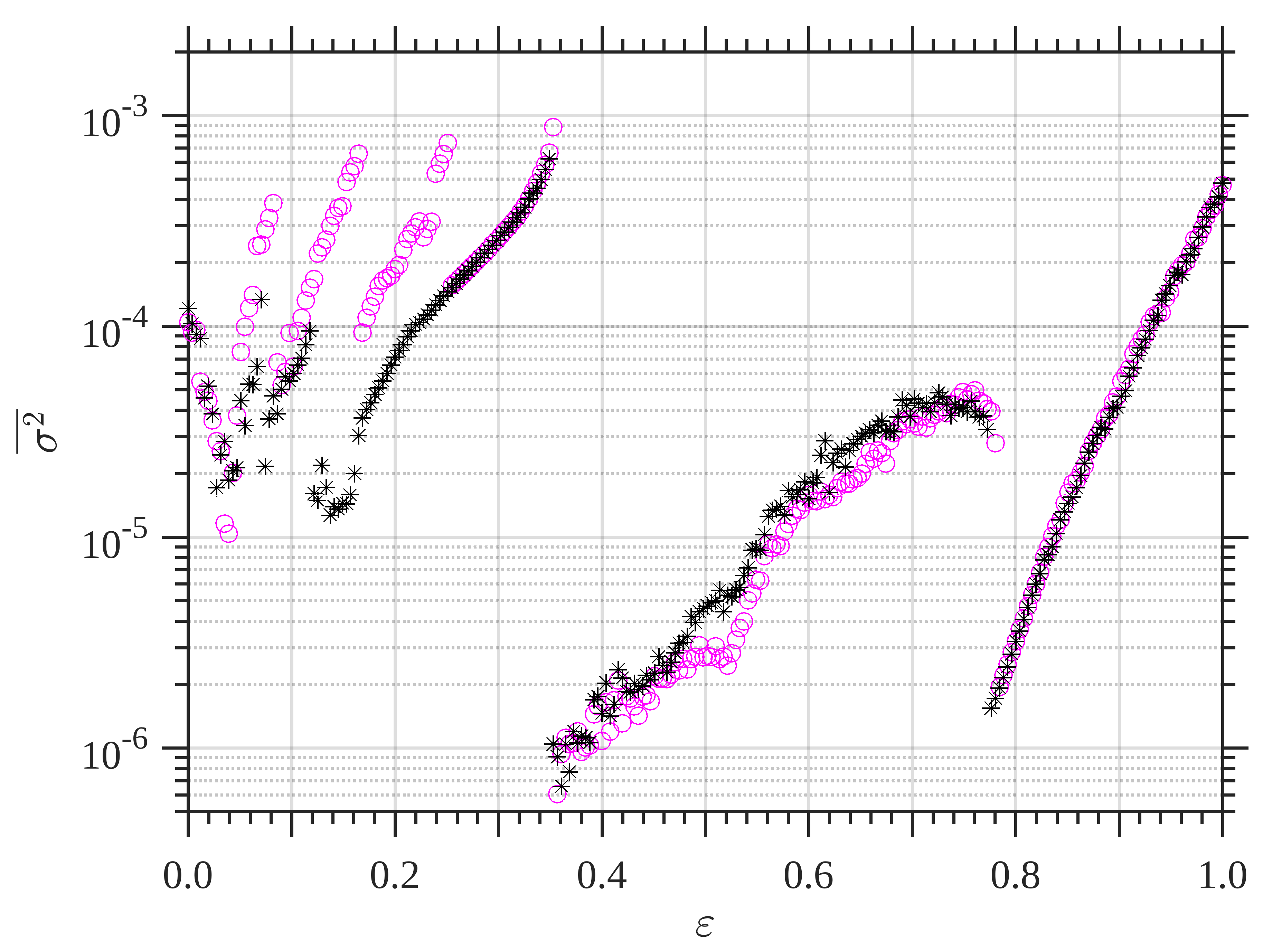} \vspace{-0.5pc}
    \caption{Hysteresis (bottom panel) in different configurations for $6$ nearly-identical, coupled, logistic maps. Parameters, $\overline{\sigma^2}$ values (top left configuration), symbols, and colours are as in Figs.~\ref{fig_1BranchNets}.}
    \label{fig_4BranchNets}
\end{figure}

\begin{figure}[htbp]
    \centering
    \includegraphics[width=0.32\columnwidth]{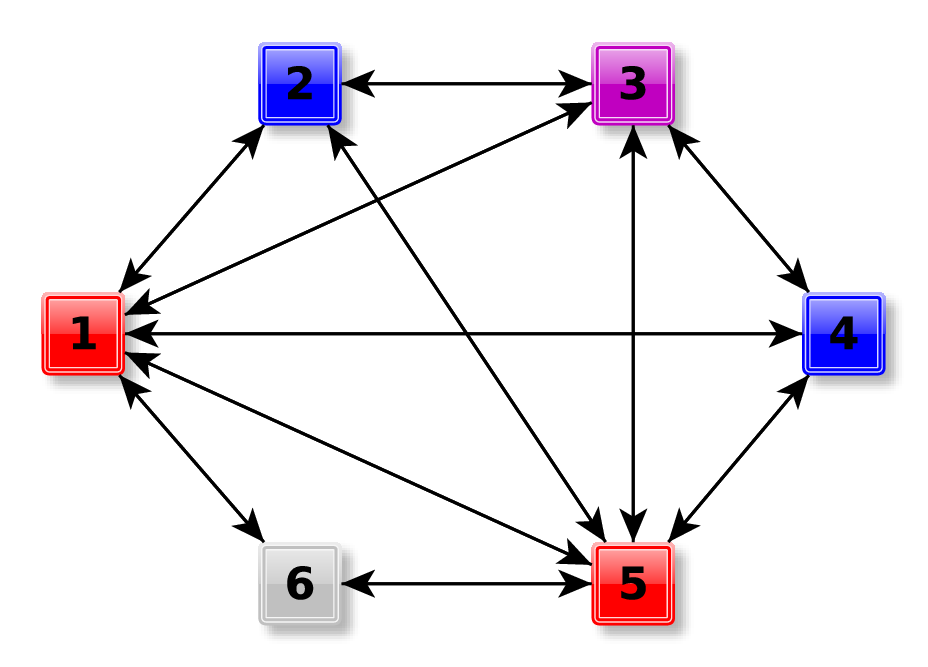}
    \includegraphics[width=0.32\columnwidth]{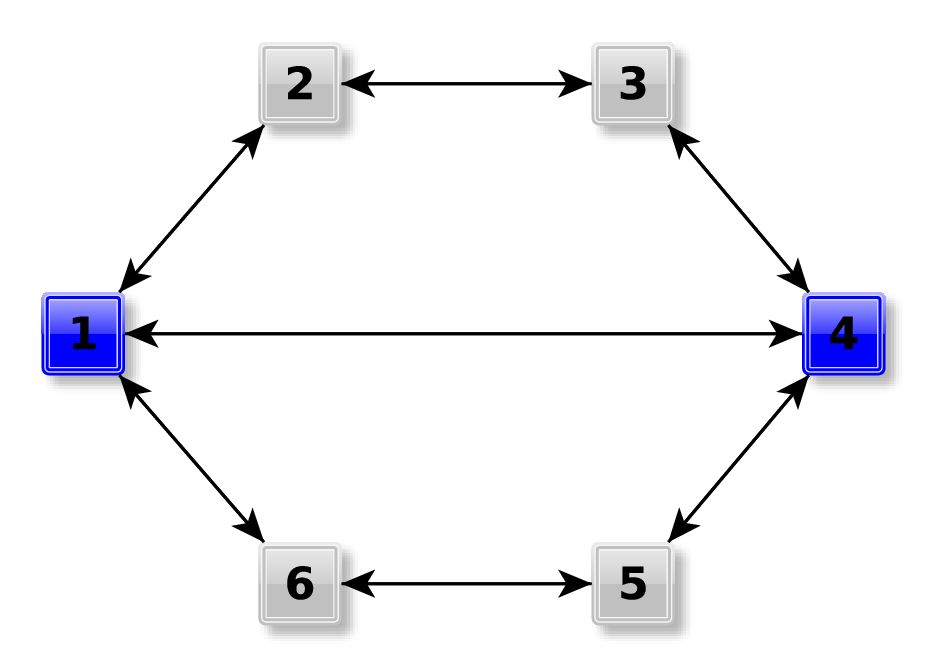}
    \includegraphics[width=0.32\columnwidth]{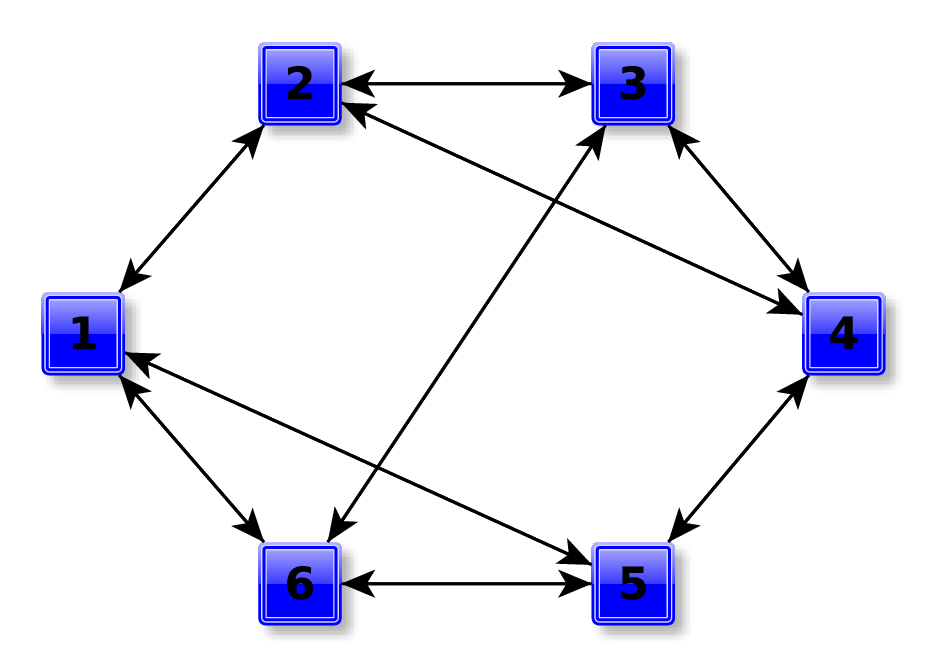}
    \includegraphics[width=0.32\columnwidth]{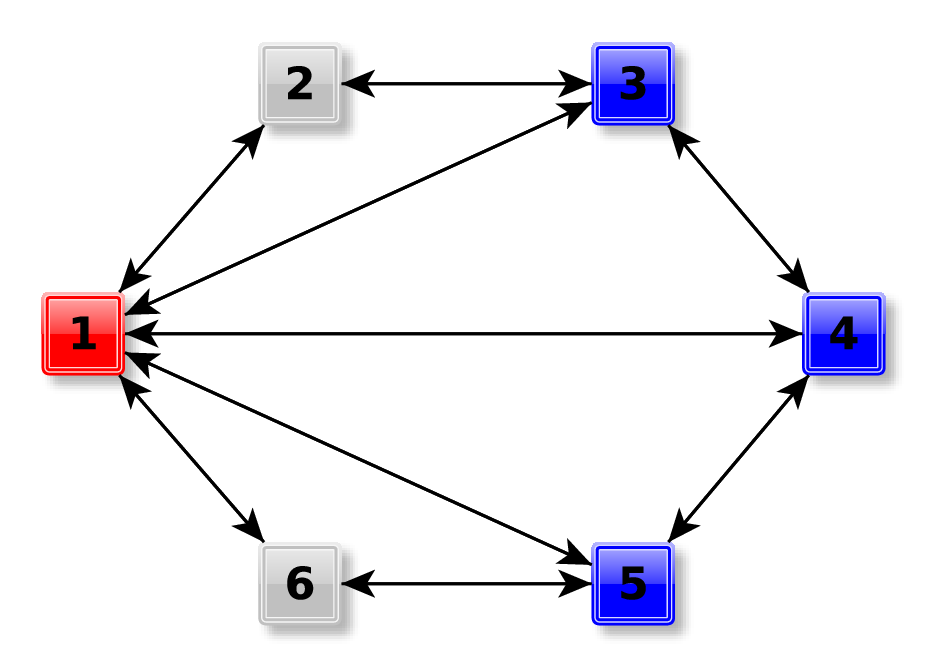}
    \includegraphics[width=0.32\columnwidth]{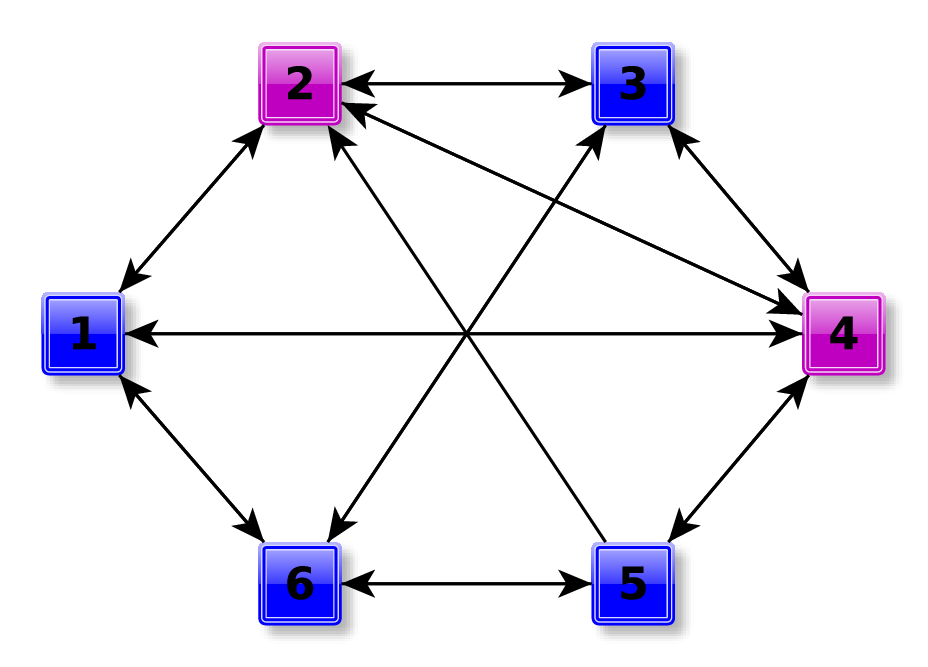}
    \includegraphics[width=0.32\columnwidth]{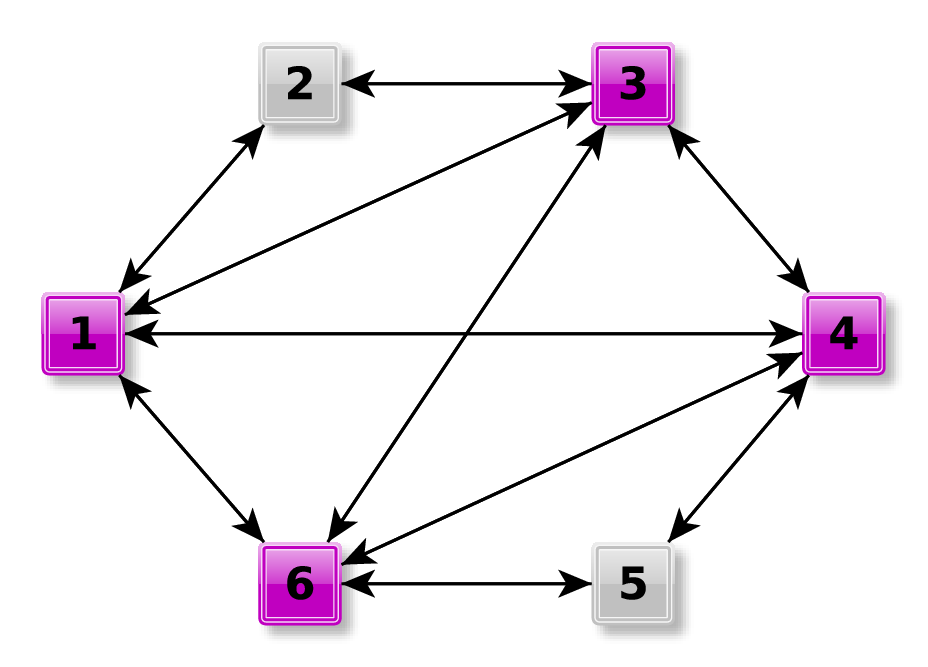}
    \includegraphics[width=0.32\columnwidth]{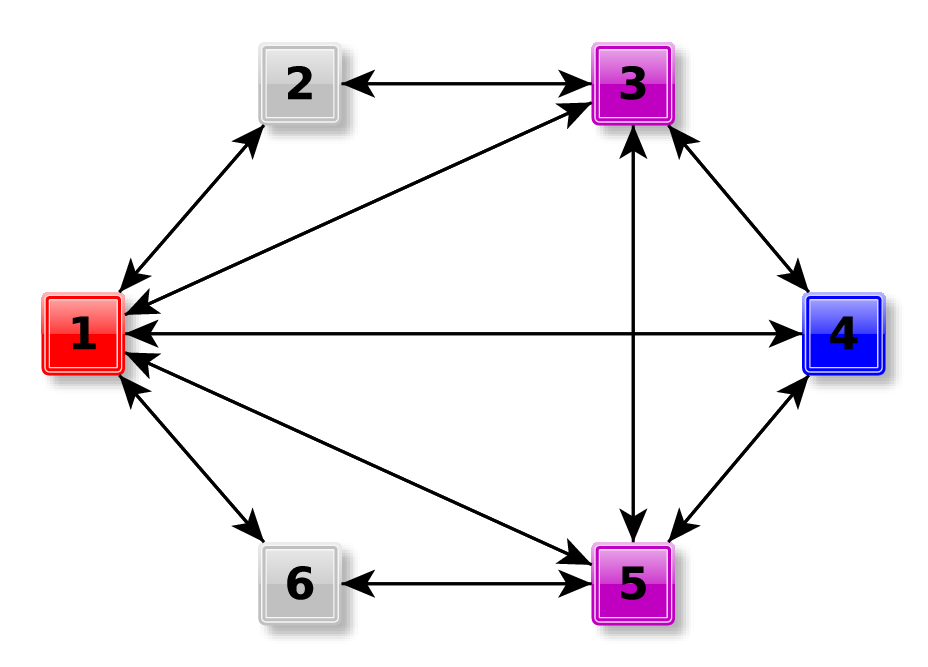}
    \includegraphics[width=0.32\columnwidth]{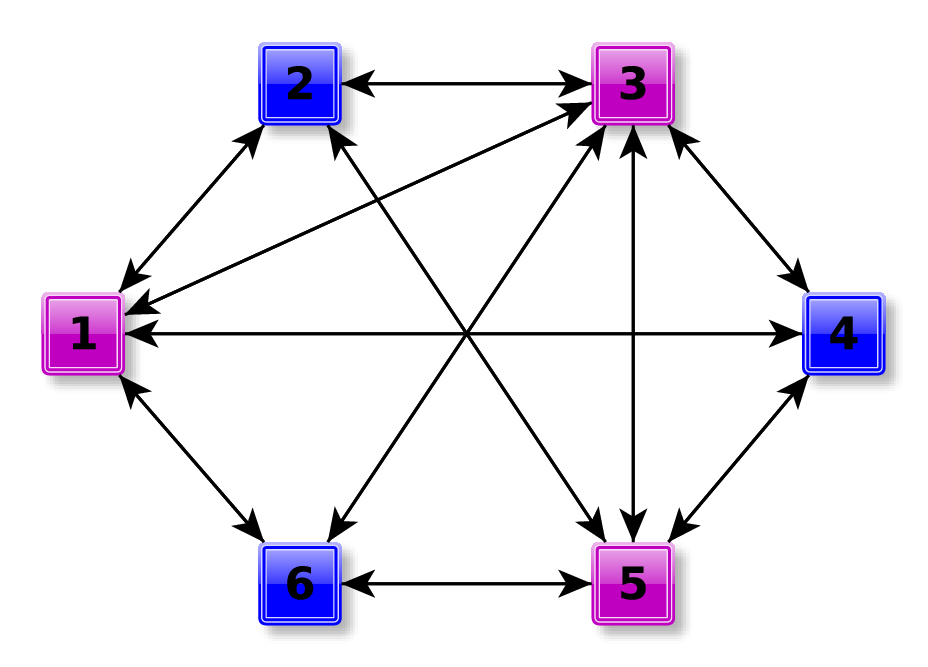}
    \includegraphics[width=0.32\columnwidth]{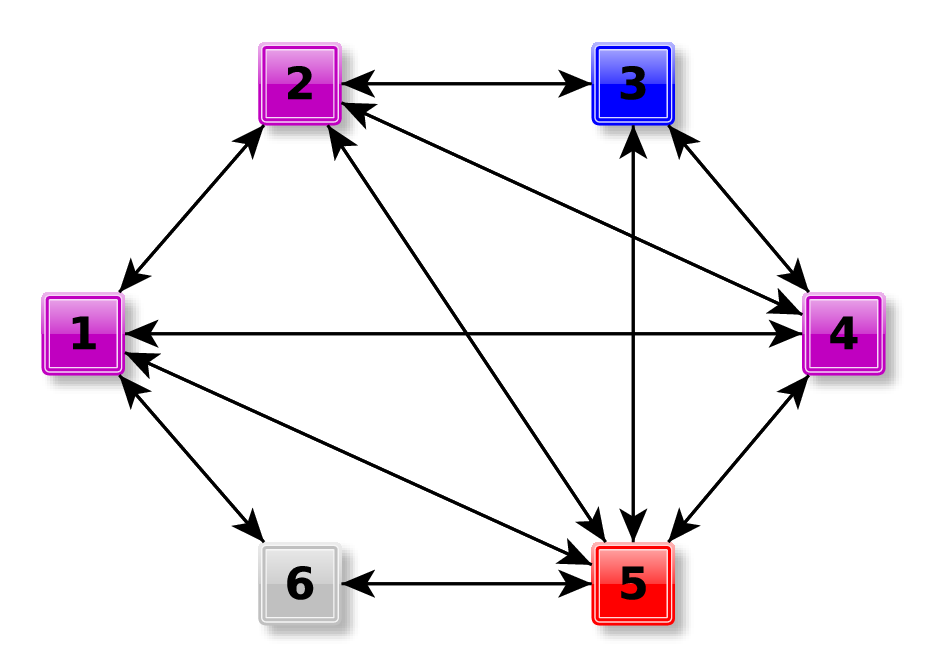}
    \includegraphics[width=0.32\columnwidth]{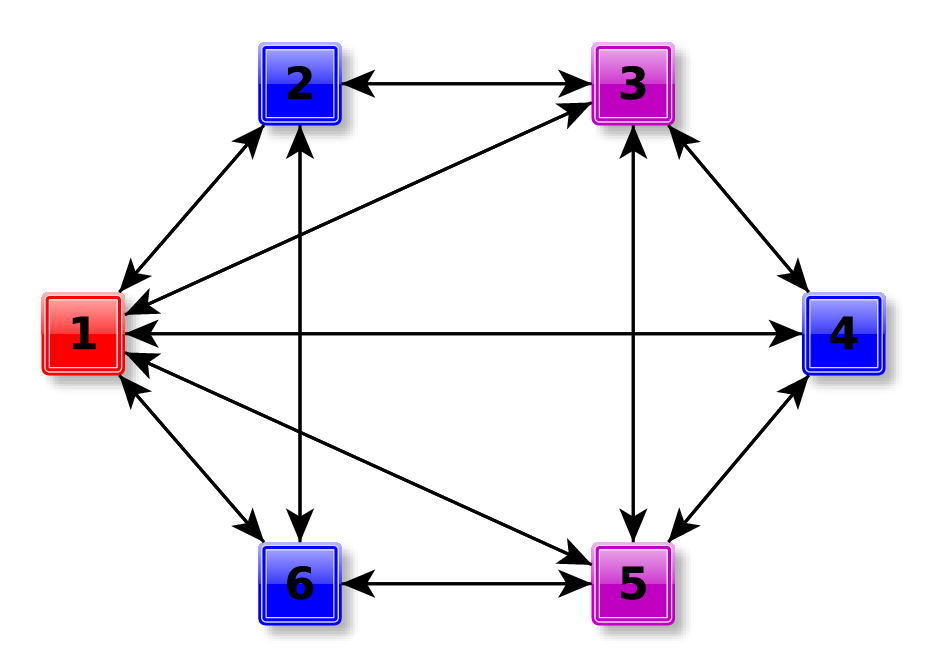}
    \includegraphics[width=0.32\columnwidth]{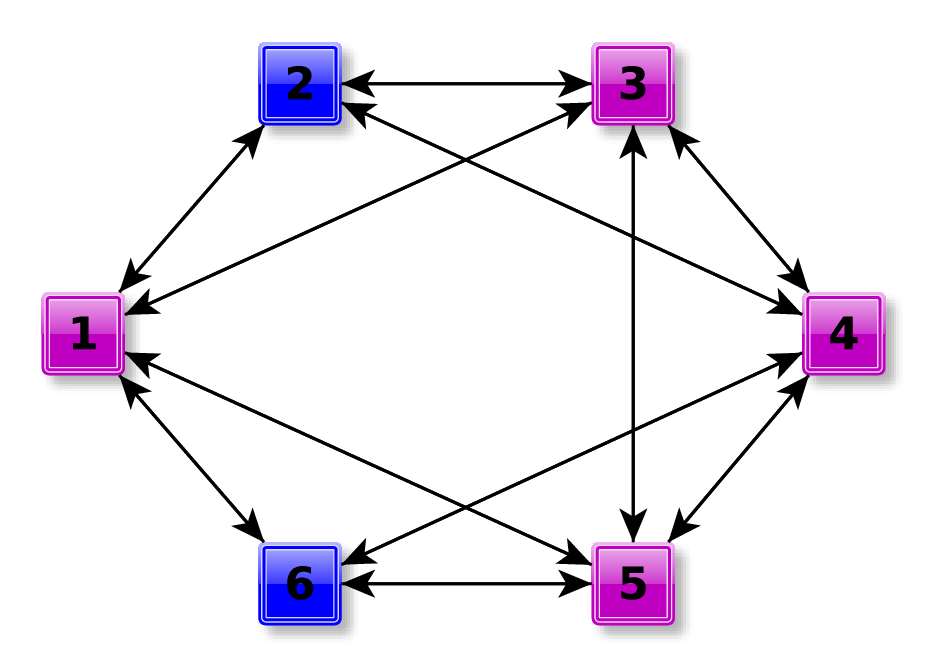}
    \includegraphics[width=0.32\columnwidth]{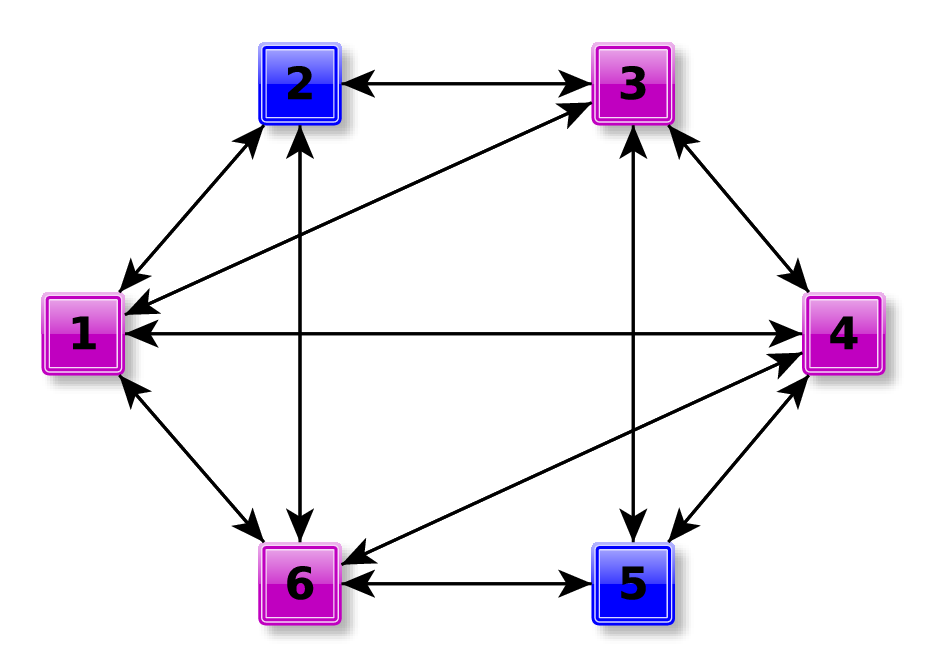}
    \includegraphics[width=0.32\columnwidth]{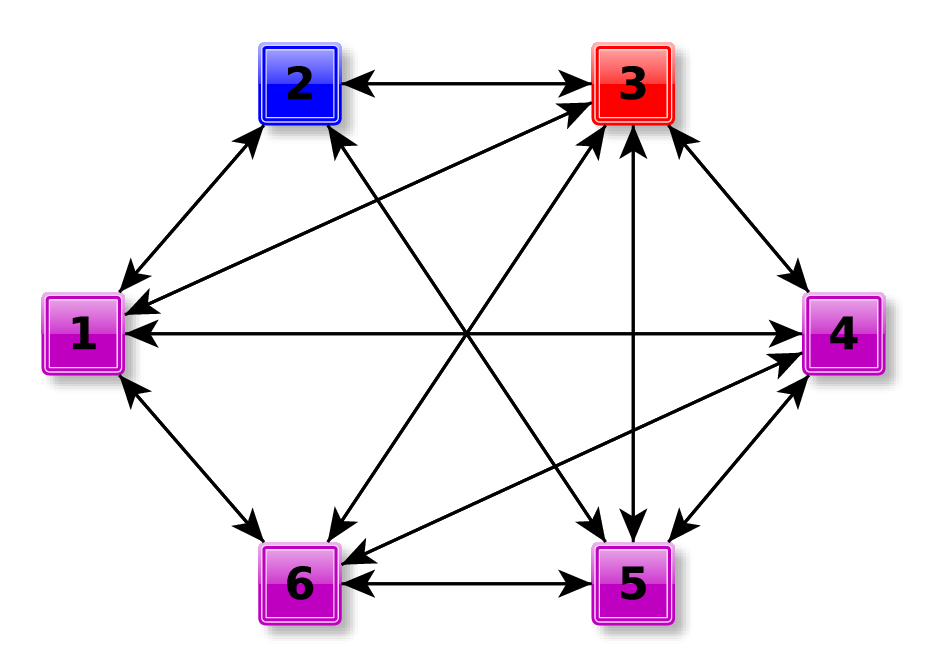}
    \includegraphics[width=0.32\columnwidth]{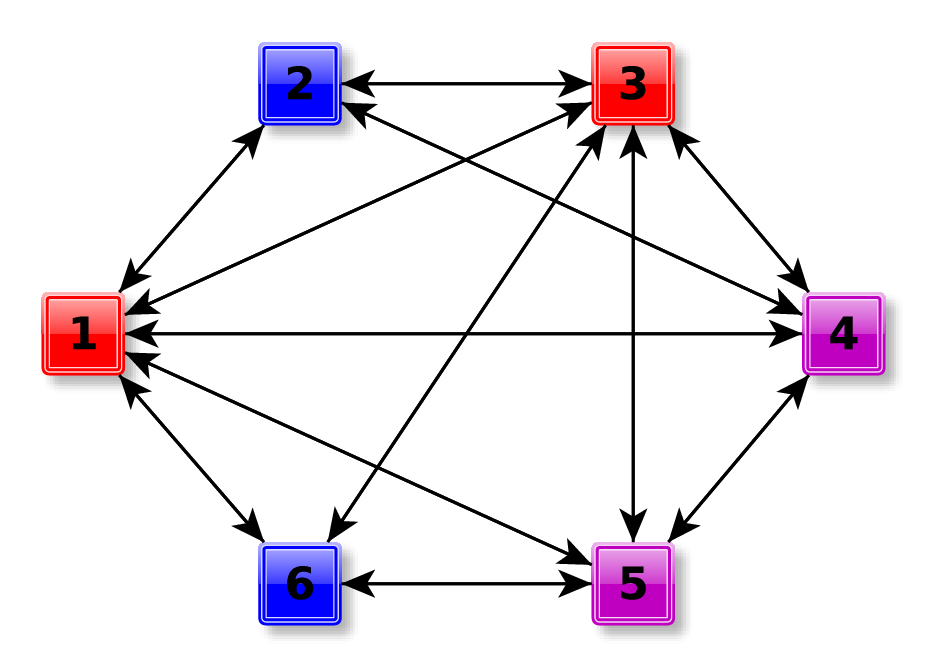}
    \includegraphics[width=0.32\columnwidth]{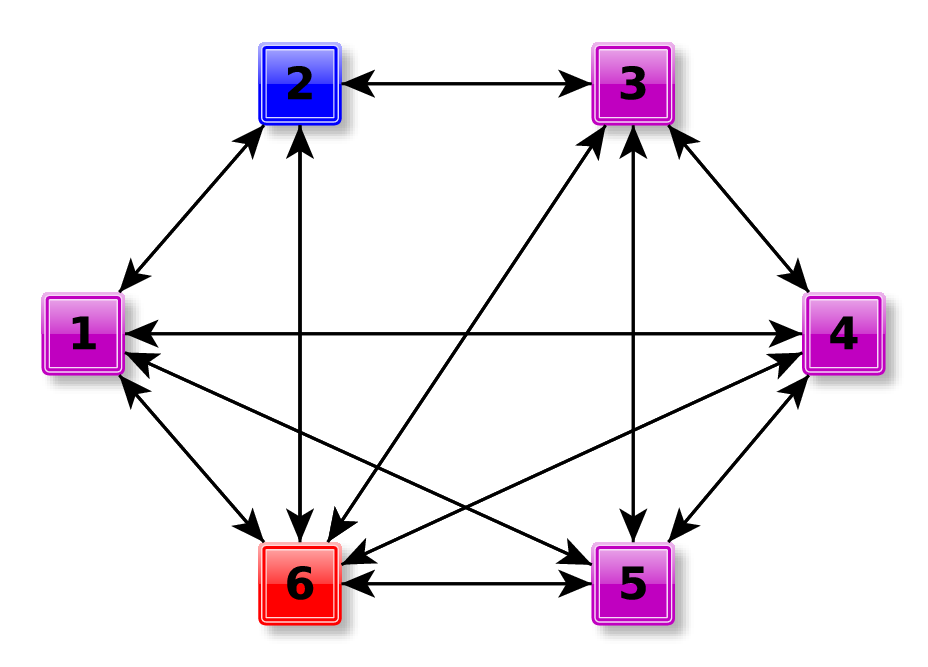}
    \includegraphics[width=0.32\columnwidth]{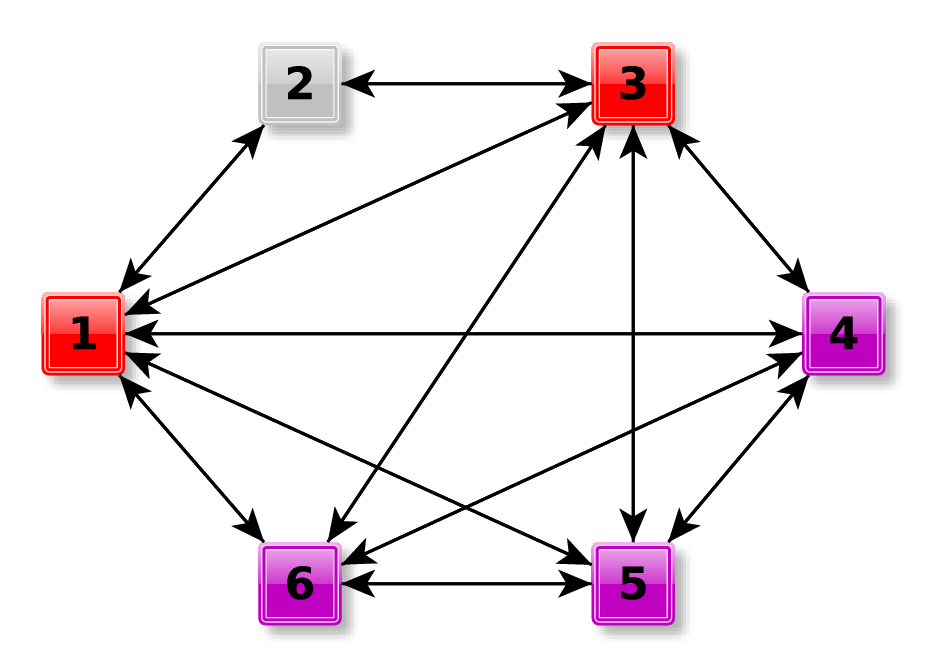}
    \includegraphics[width=0.32\columnwidth]{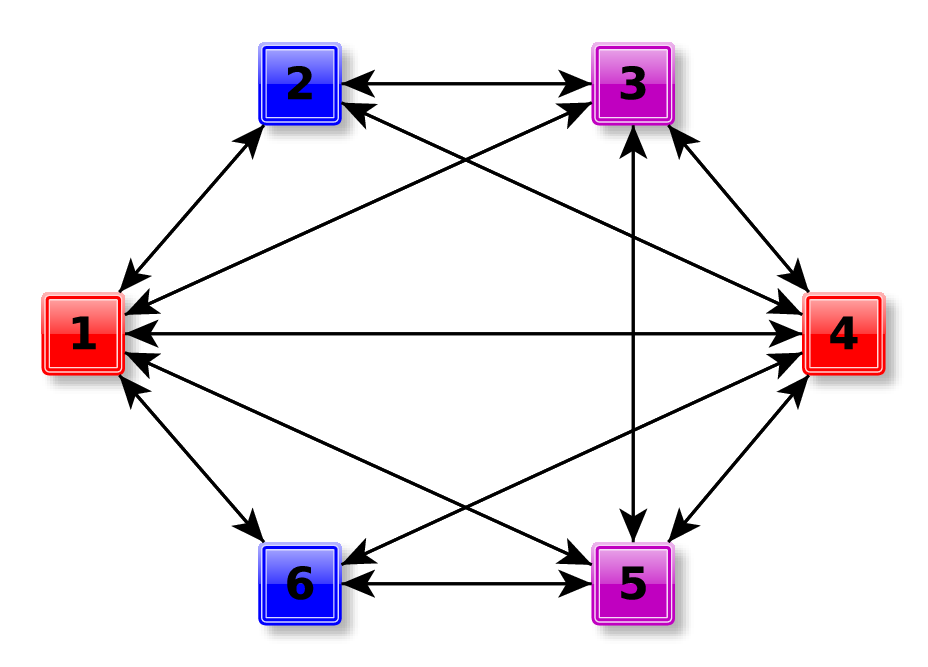}
    \includegraphics[width=0.32\columnwidth]{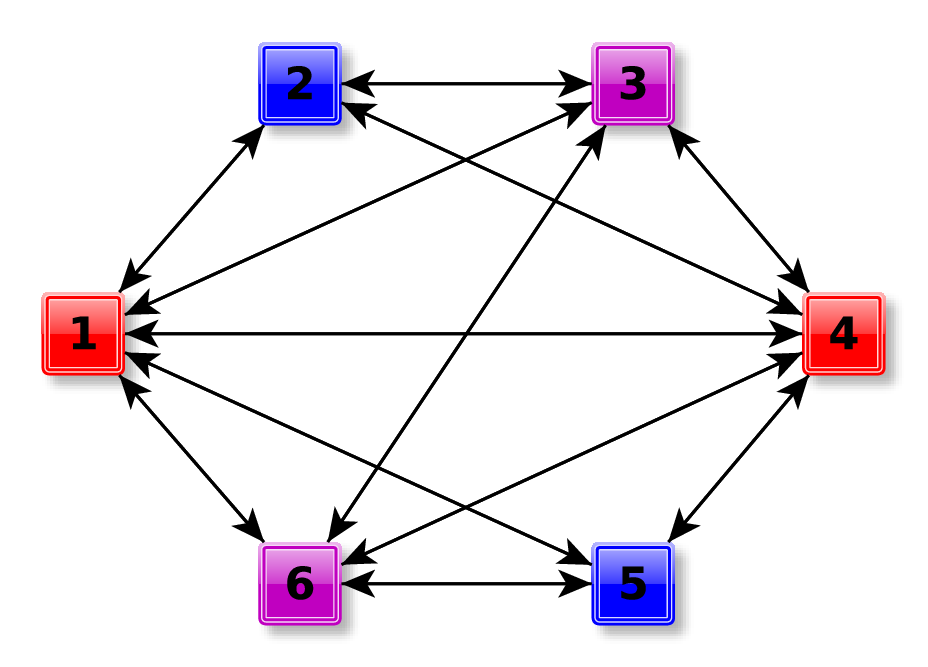}
    \includegraphics[width=0.32\columnwidth]{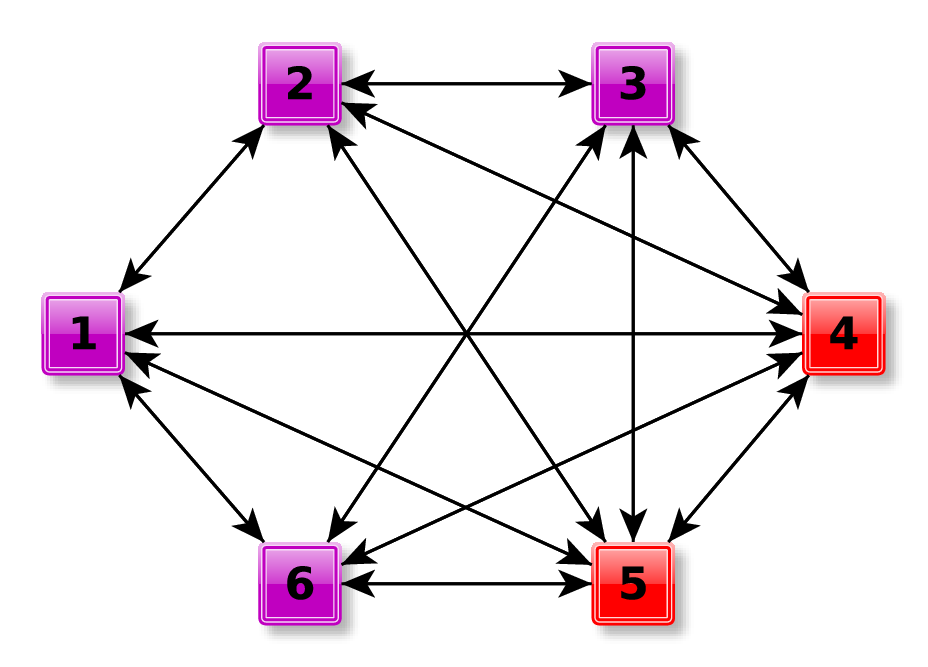}
    \includegraphics[width=0.32\columnwidth]{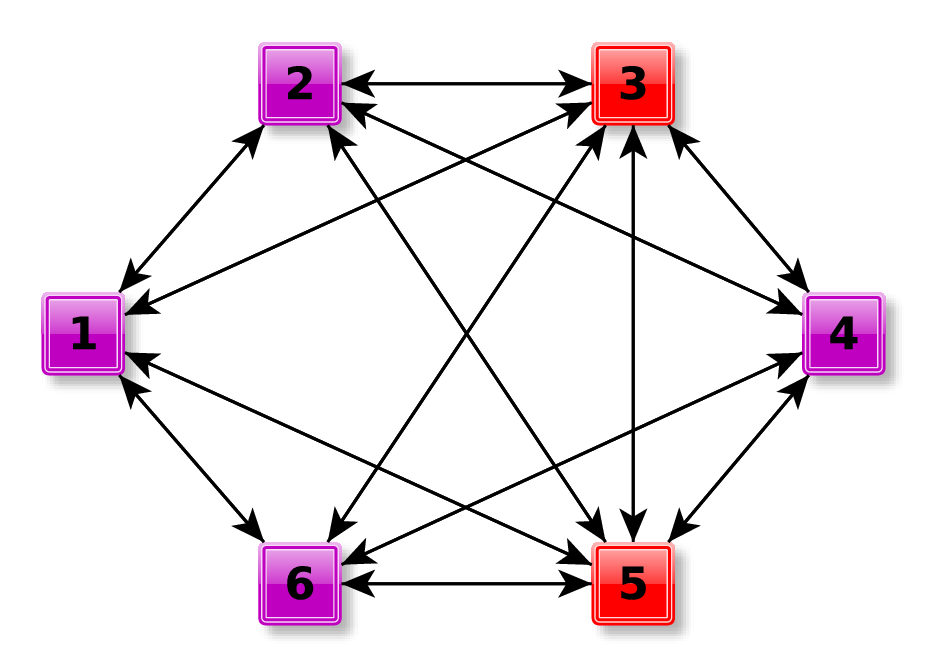}
    \includegraphics[width=0.32\columnwidth]{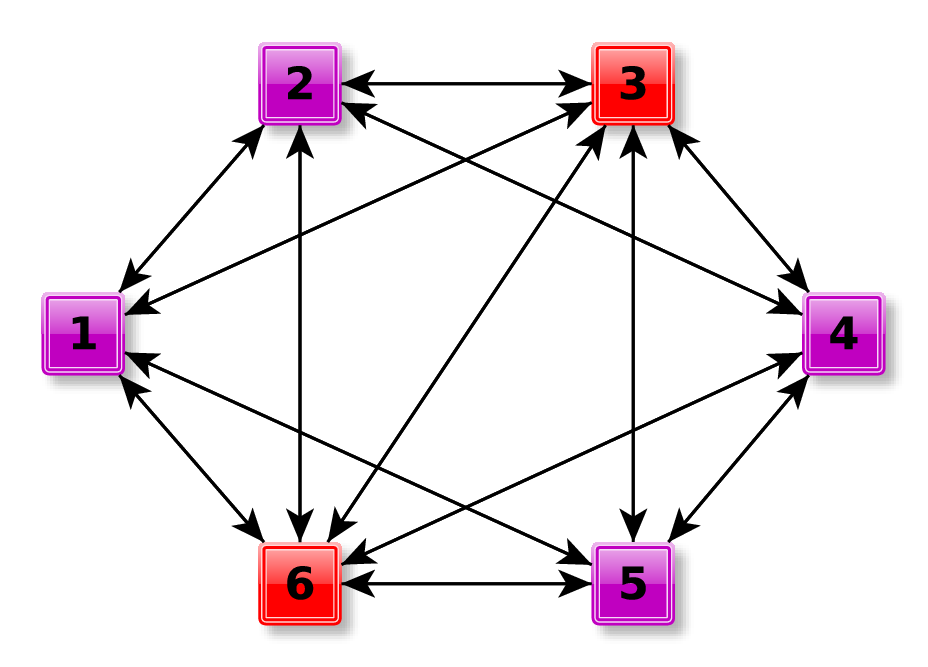}
    \includegraphics[width=0.32\columnwidth]{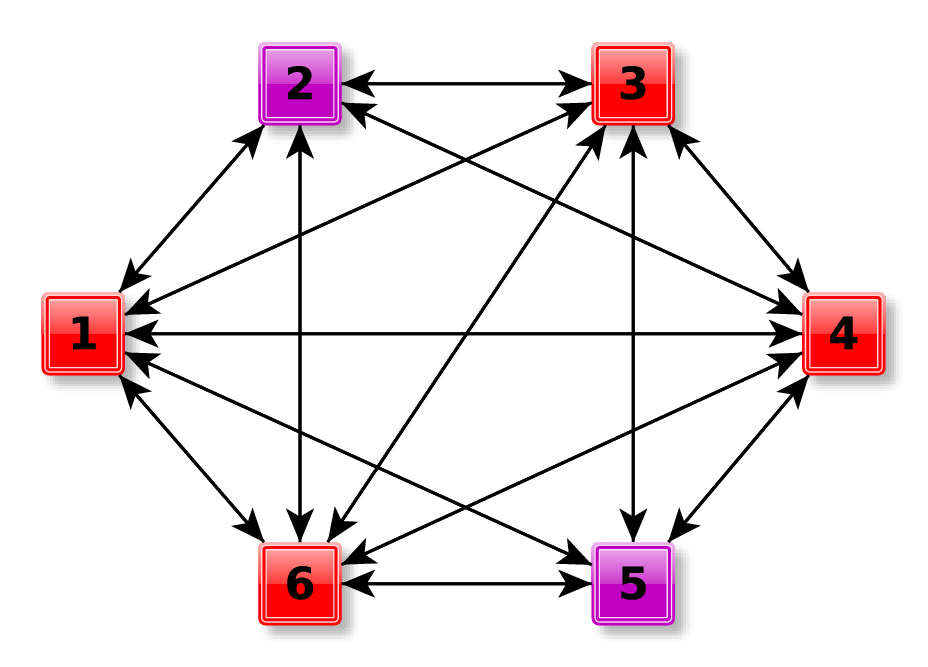}
    \includegraphics[width=0.95\columnwidth]{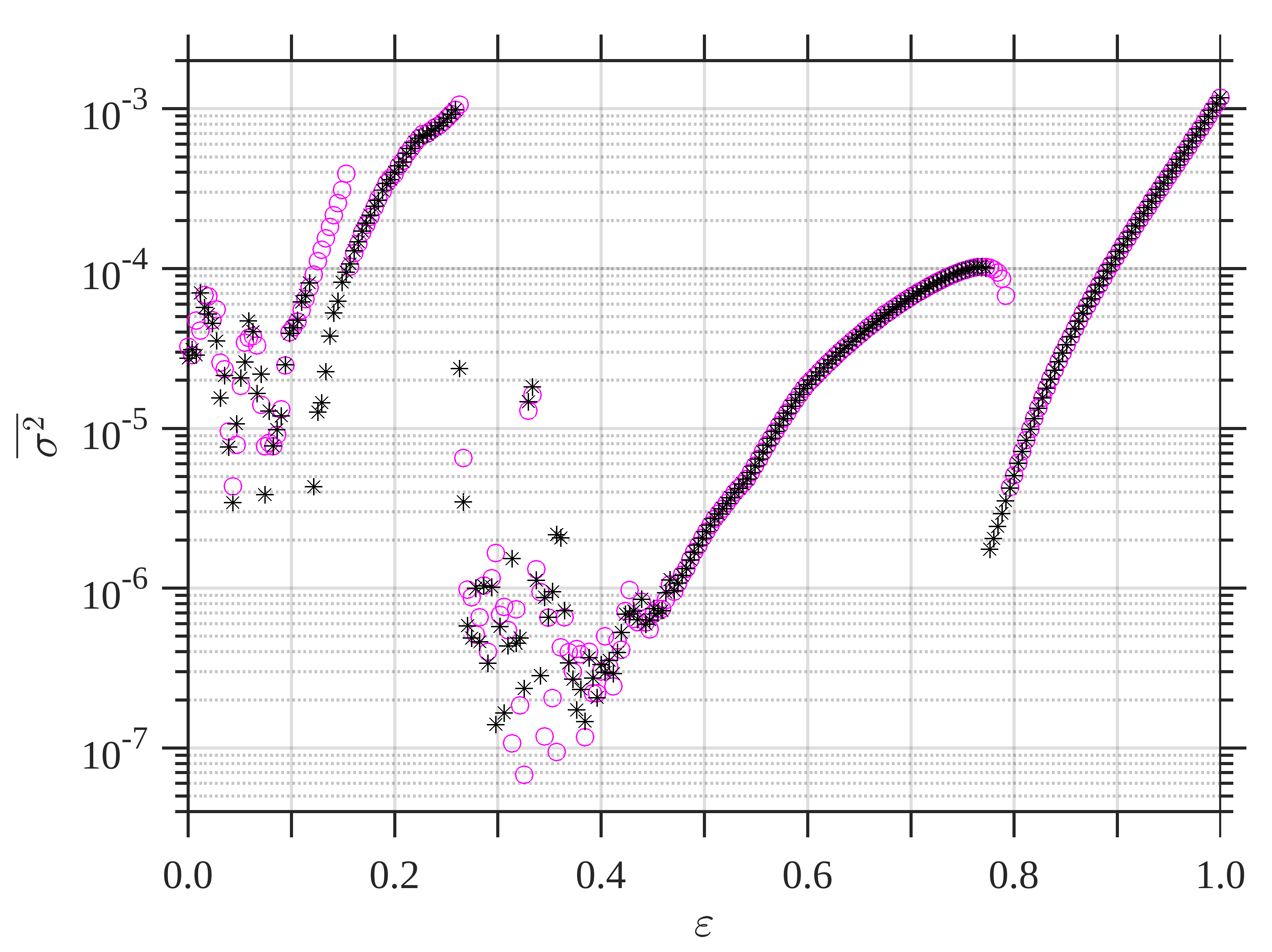} \vspace{-0.5pc}
    \caption{Hysteresis (bottom panel) in different configurations for $6$ nearly-identical ($r_i \simeq r = 3.75\pm0.03\,\forall\,i$), coupled, logistic maps. Parameters, symbols, and colours are as in Figs.~\ref{fig_1BranchNets}. Particularly, these $\overline{\sigma^2}$ values correspond to the top left coupling-configuration.}
    \label{fig_2BranchNets}
\end{figure}

Figures~\ref{fig_2BranchNets} and \ref{fig_3BranchNets} show $2$ and $3$ branches, respectively, for $\overline{\sigma^2}$ as $\varepsilon$ is increased (magenta circles); which group most of the $47$ configurations. In particular, the order parameter in Fig.~\ref{fig_2BranchNets} (bottom panel) seems to be the predominant behaviour, where we find $22$ configurations showing this collective behaviour and $17$ configurations showing the behaviour in Fig.~\ref{fig_3BranchNets} (bottom panel).

\begin{figure}[h!]
    \centering
    \includegraphics[width=0.32\columnwidth]{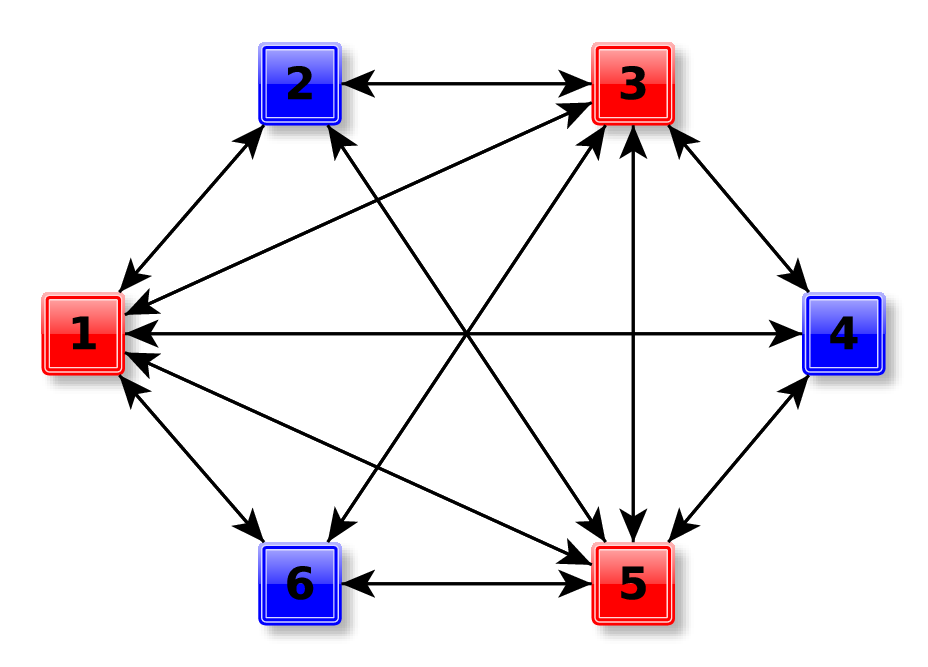}
    \includegraphics[width=0.32\columnwidth]{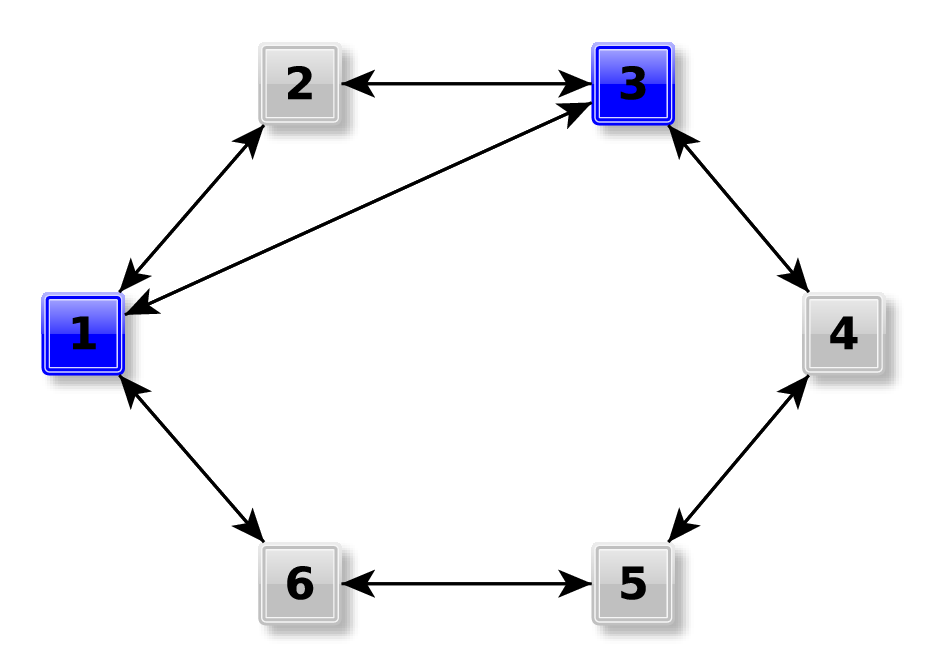}
    \includegraphics[width=0.32\columnwidth]{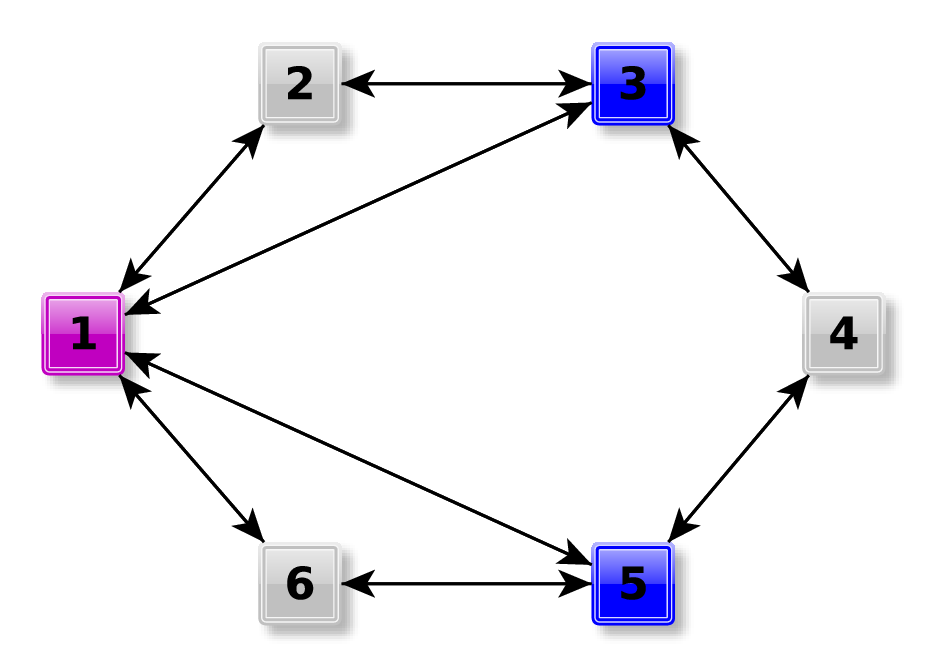}
    \includegraphics[width=0.32\columnwidth]{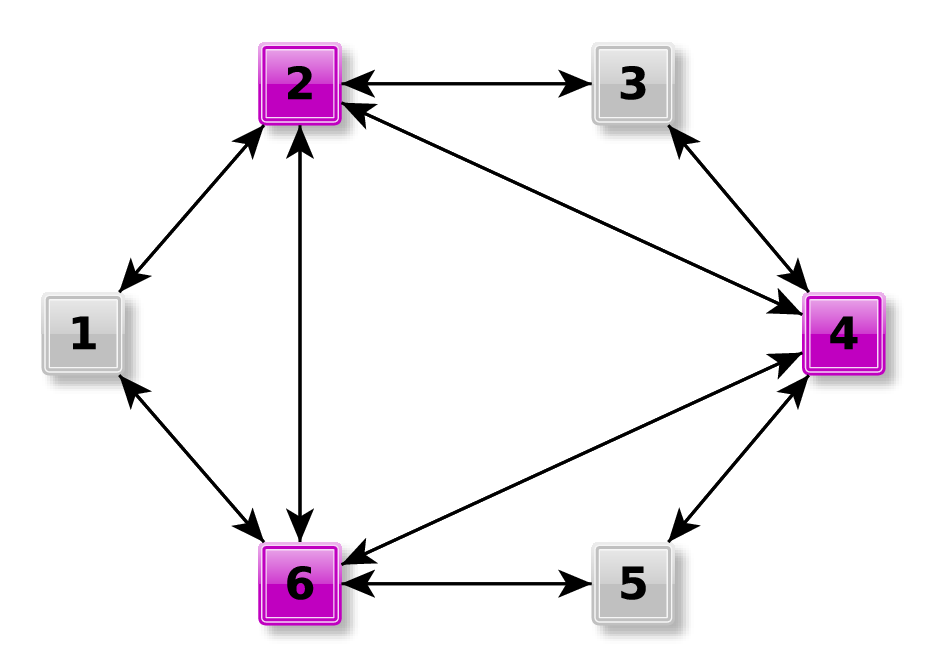}
    \includegraphics[width=0.32\columnwidth]{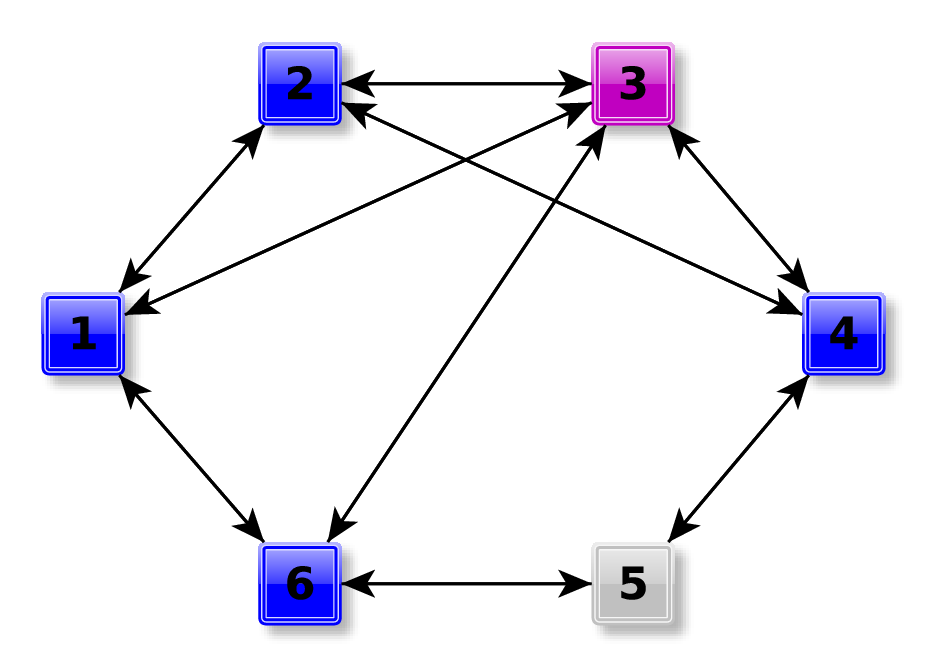}
    \includegraphics[width=0.32\columnwidth]{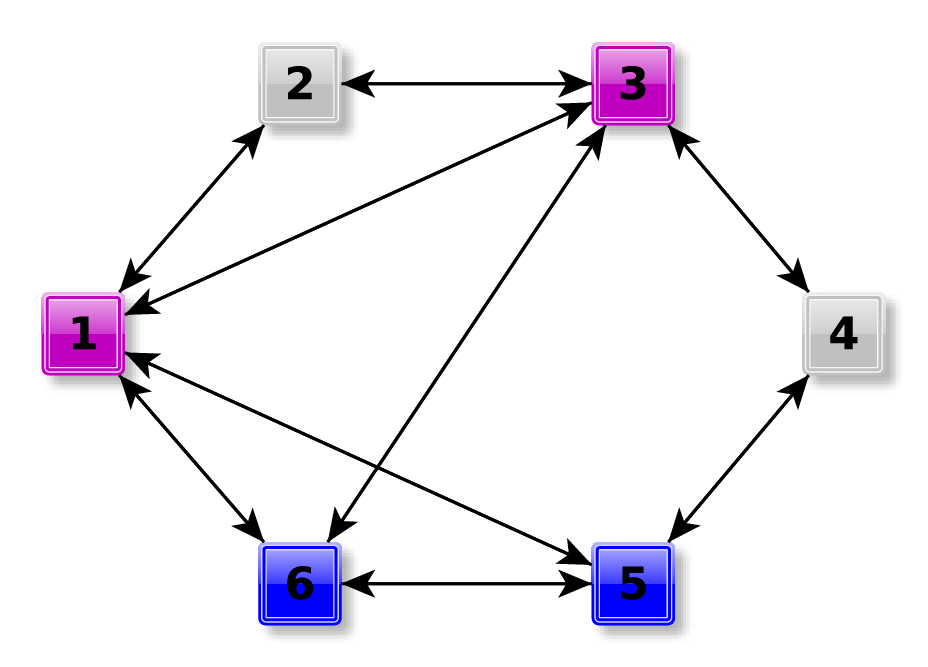}
    \includegraphics[width=0.32\columnwidth]{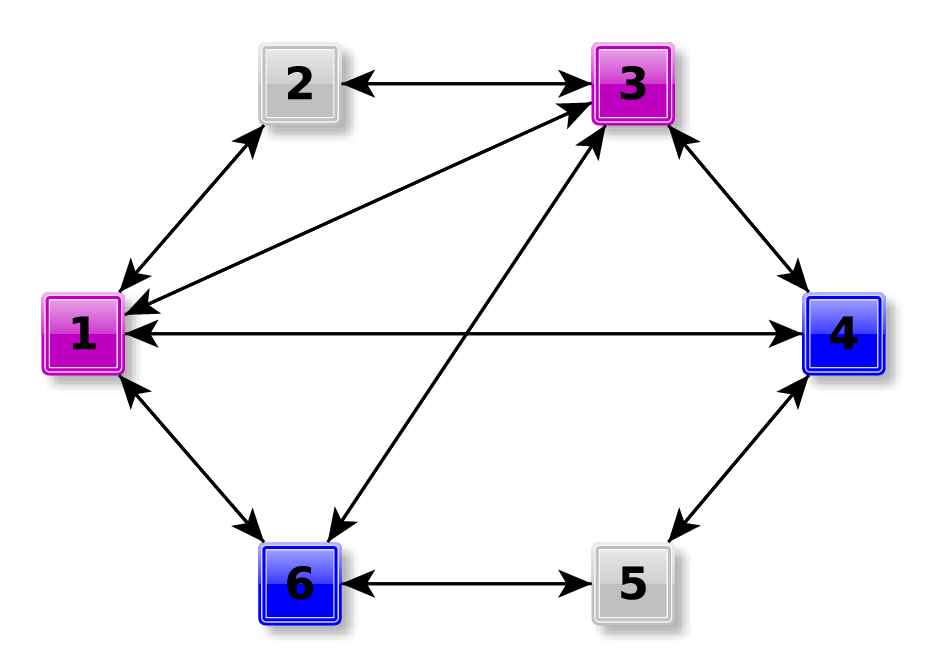}
    \includegraphics[width=0.32\columnwidth]{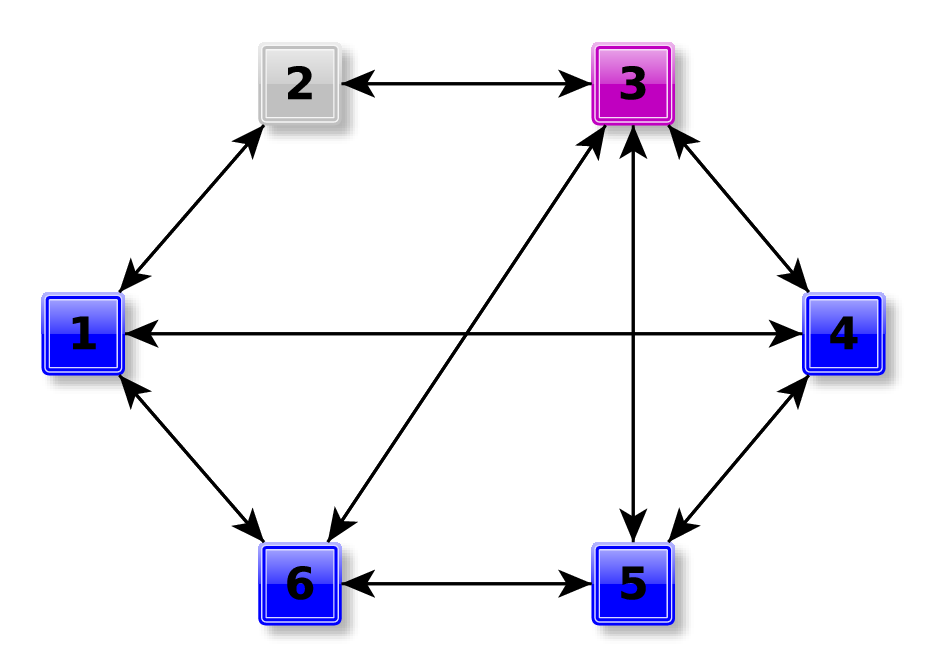}
    \includegraphics[width=0.32\columnwidth]{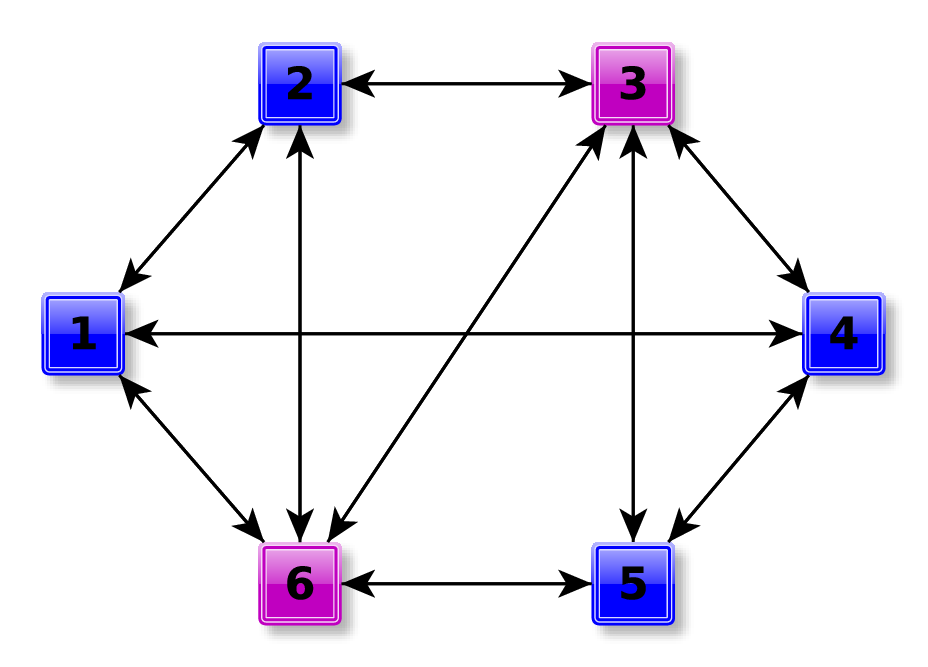}
    \includegraphics[width=0.32\columnwidth]{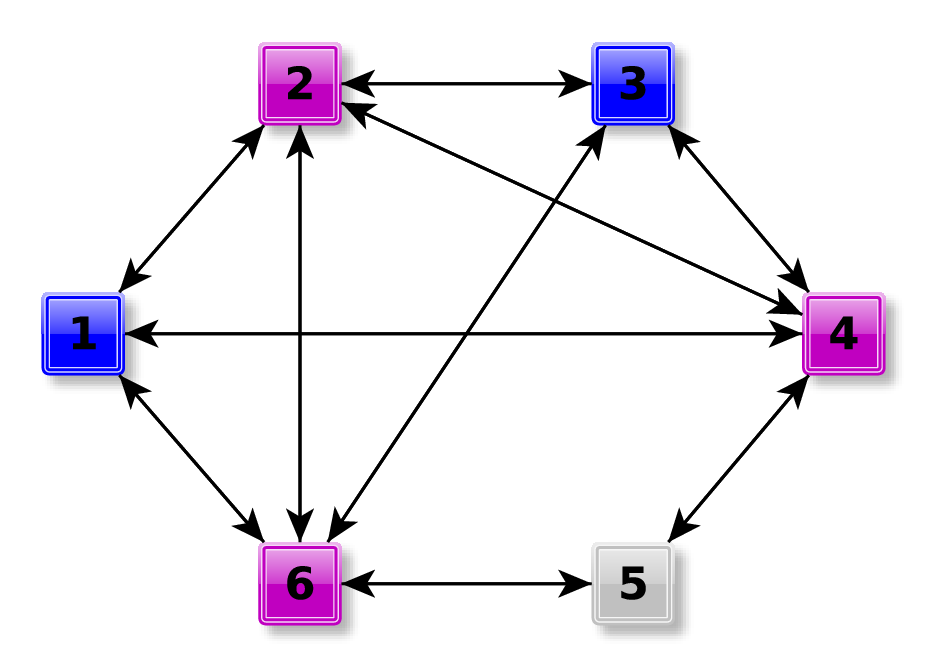}
    \includegraphics[width=0.32\columnwidth]{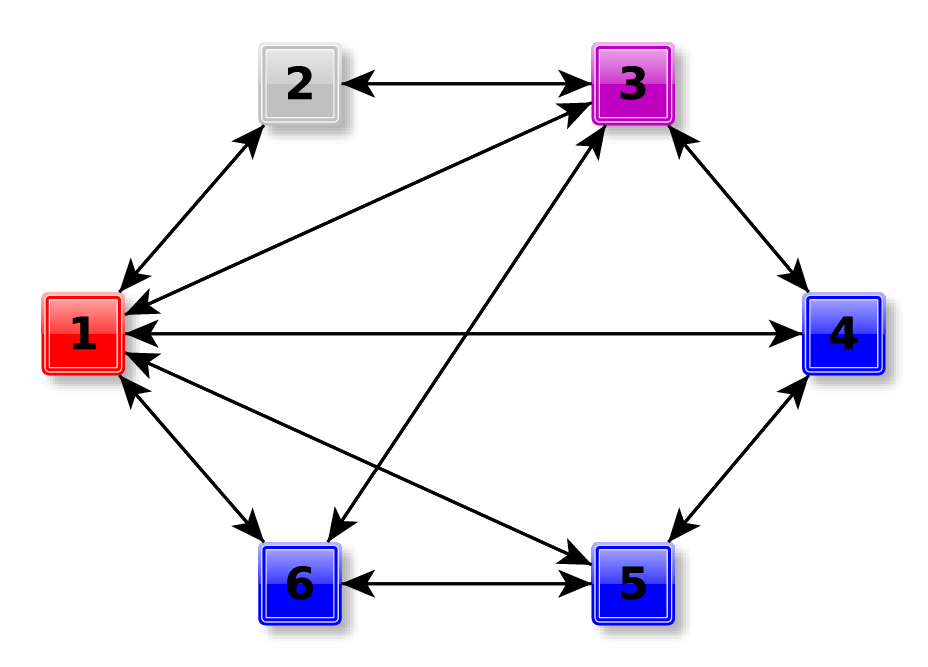}
    \includegraphics[width=0.32\columnwidth]{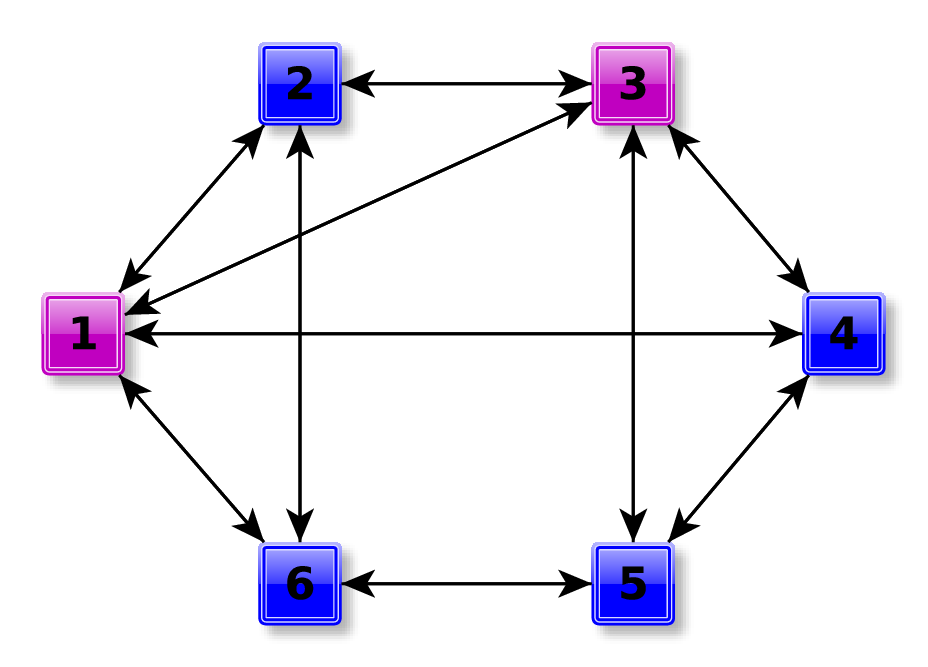}
    \includegraphics[width=0.32\columnwidth]{25.png}
    \includegraphics[width=0.32\columnwidth]{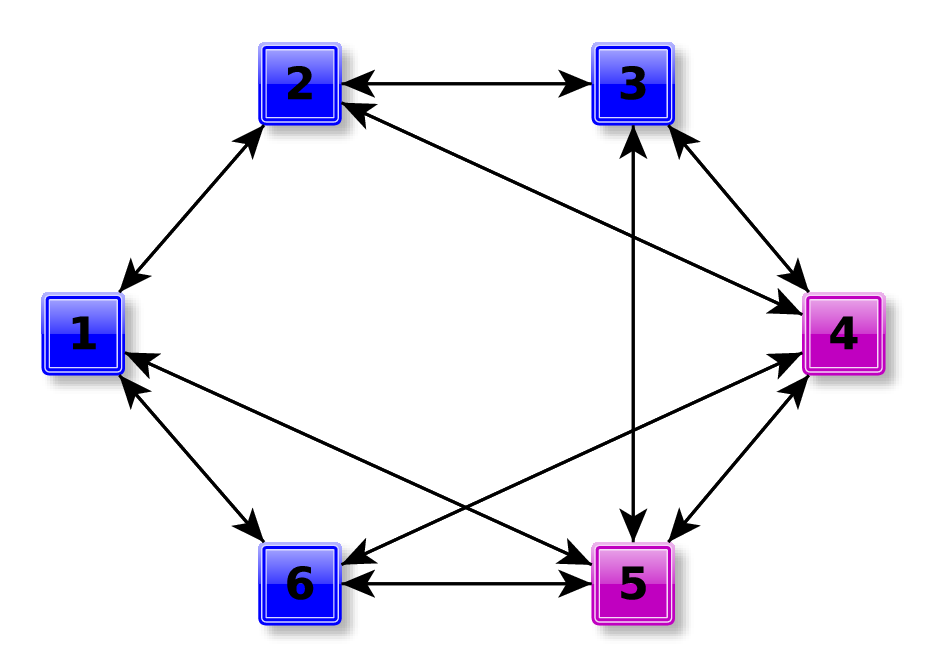}
    \includegraphics[width=0.32\columnwidth]{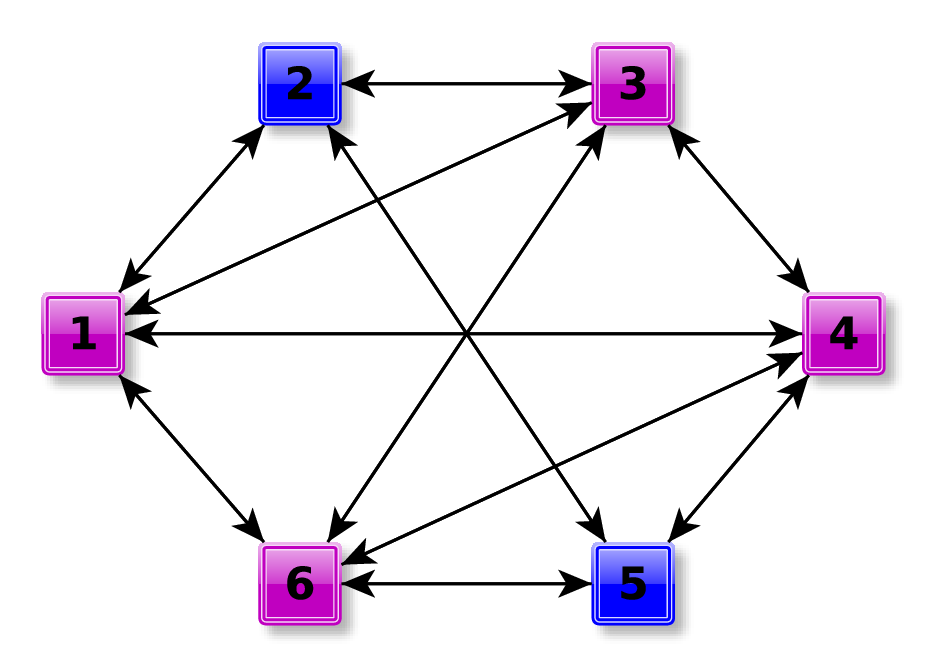}
    \includegraphics[width=0.32\columnwidth]{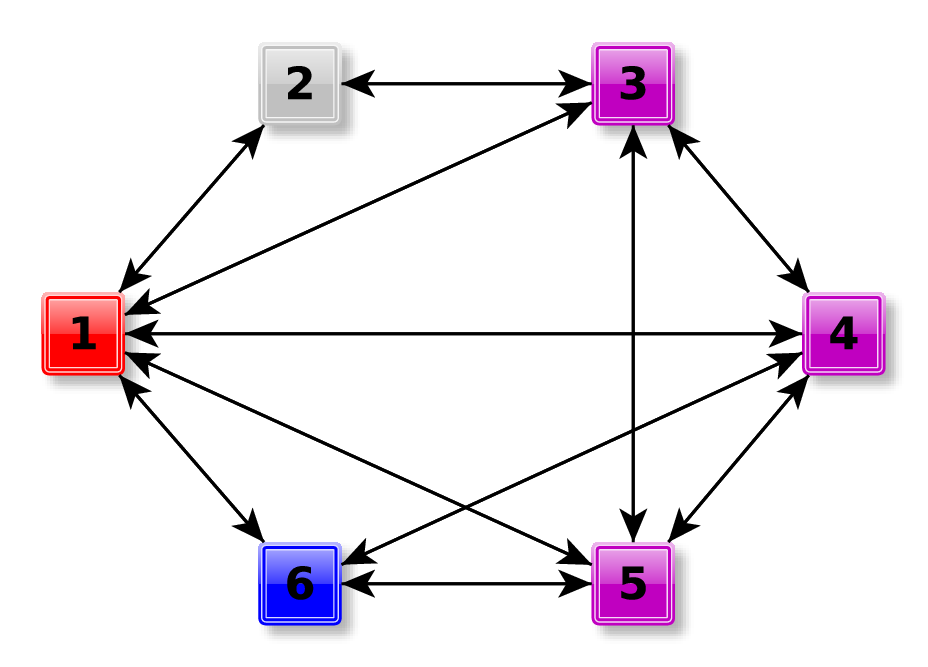}
    \includegraphics[width=0.32\columnwidth]{36.png}
    \includegraphics[width=0.32\columnwidth]{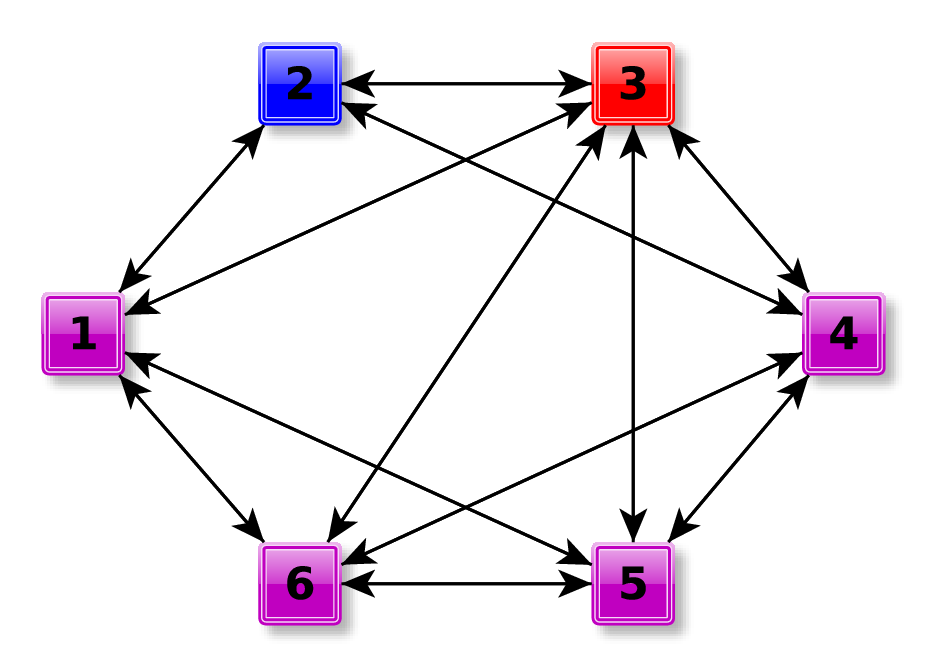}
    \includegraphics[width=0.95\columnwidth]{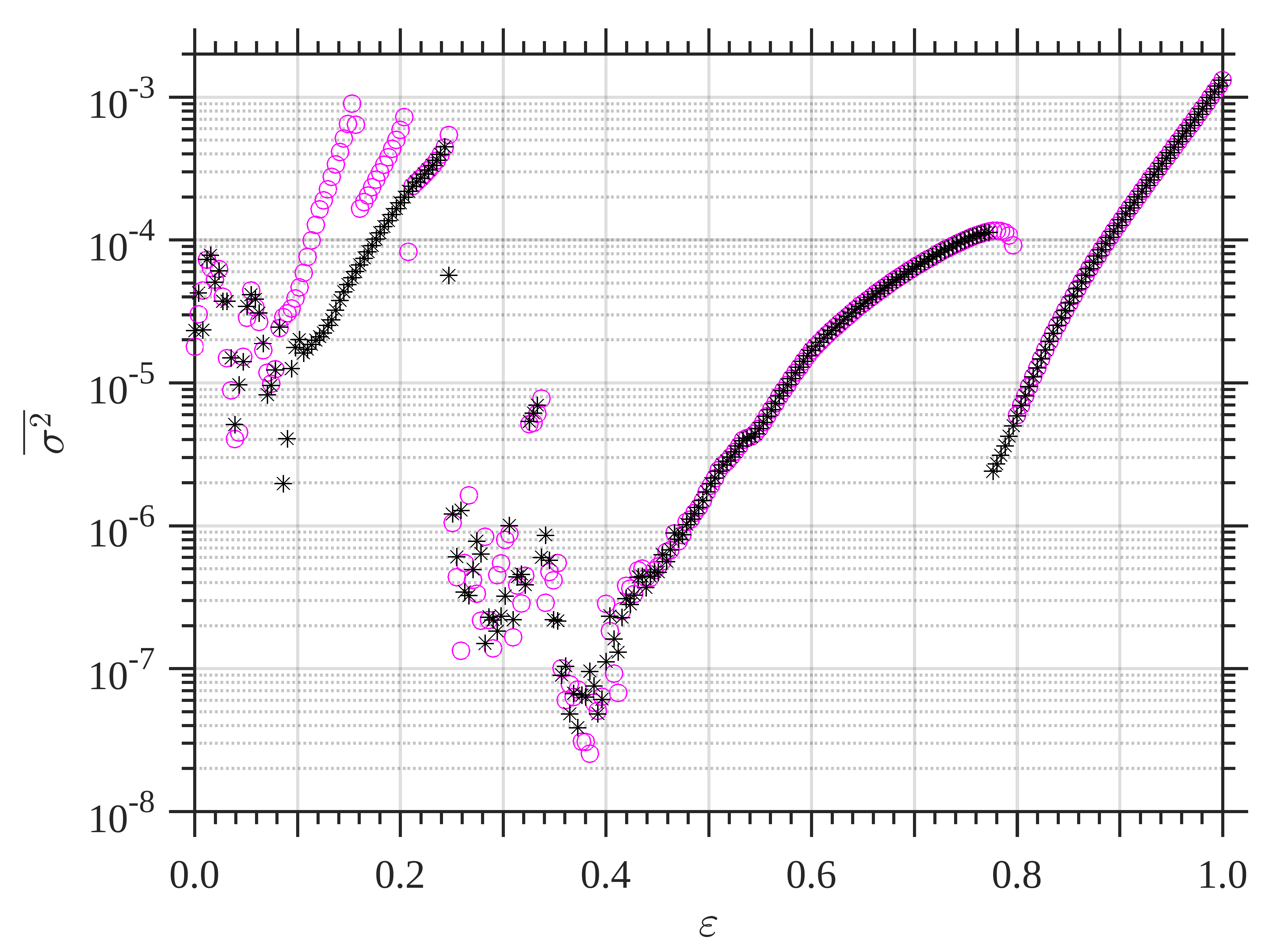} \vspace{-0.5pc}
    \caption{Hysteresis (bottom panel) in different configurations for $6$ nearly-identical ($r_i \simeq r = 3.75\pm0.03\,\forall\,i$), coupled, logistic maps. Parameters, symbols, and colours are as in Figs.~\ref{fig_1BranchNets}. Particularly, these $\overline{\sigma^2}$ values correspond to the top left coupling-configuration.}
    \label{fig_3BranchNets}
\end{figure}
%
	\subsection{Hysteresis for strong couplings} \label{sec_LargeE}
\begin{figure}[htbp]
    \centering
    \includegraphics[width=0.32\columnwidth]{29.png}
    \includegraphics[width=0.95\columnwidth]{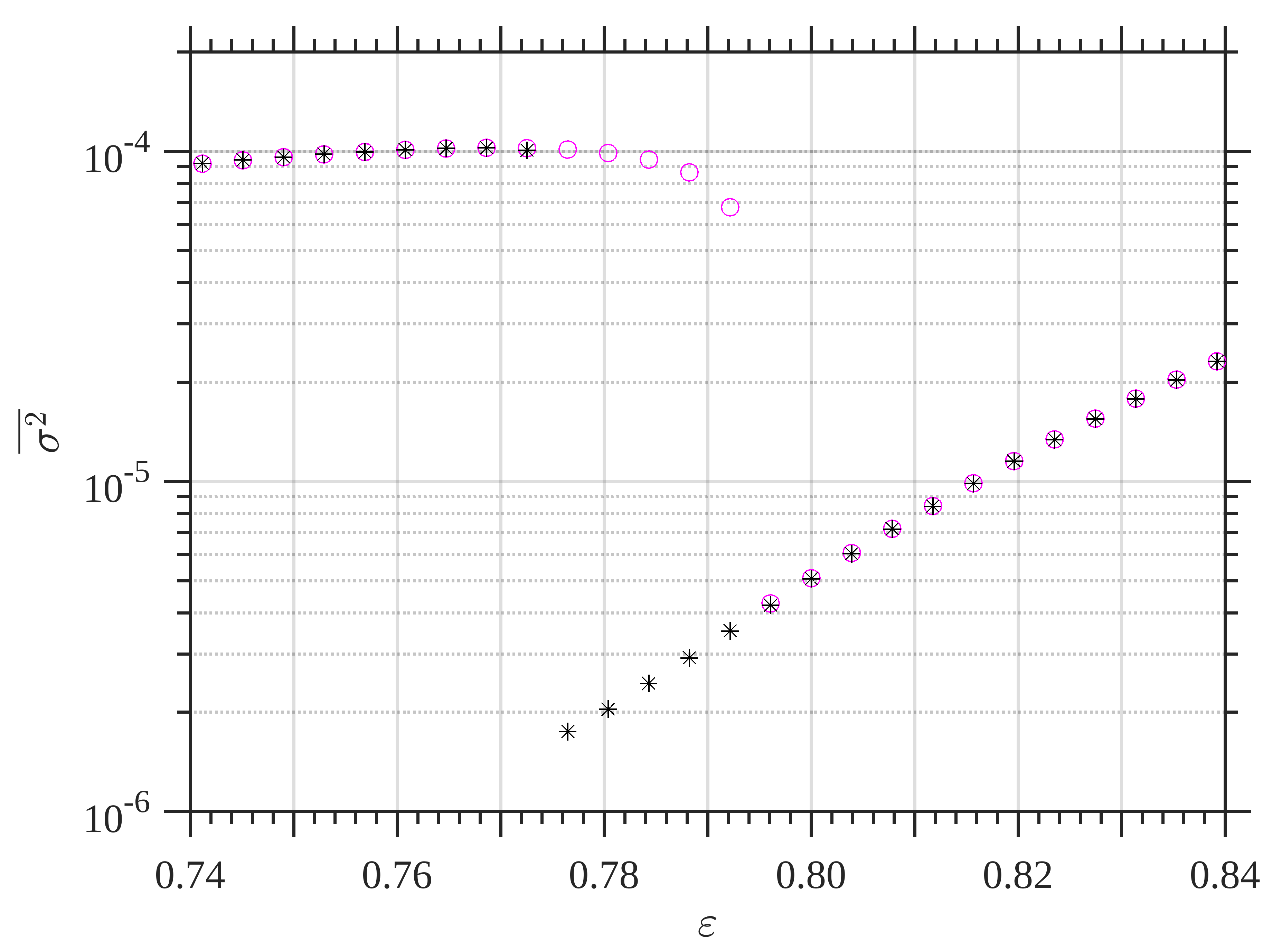} \vspace{-0.5pc}
    \caption{Hysteresis (bottom panel) for strong coupling for the first configuration in Fig.~\ref{fig_2BranchNets} (top panel). Increasing [Decreasing] coupling strength, $\varepsilon$ is shown by magenta circles [black asterisks].}
    \label{fig_LargeCouple}
\end{figure}

We note that all configurations undergo a period-doubling bifurcation for strong coupling at $\varepsilon_c \simeq 0.8$. This period-doubling bifurcation can be seen in the bifurcation diagram shown in Fig.~\ref{fig_BifExample} for $1$ of the $6$ logistic maps in the ring configuration. More importantly, we find that around this critical coupling there is another hysterical region, where the system collapses from a period $2$ attractor to a fixed-point differently when $\varepsilon$ is increased than when it is decreased. This hysteresis can be seen in the bottom panel of Fig.~\ref{fig_LargeCouple}, where the order parameter, $\overline{\sigma^2}$, shows the collective dynamics for these $2$ attractors (period $2$ orbit and fixed point) as $\varepsilon$ is increased (magenta circles) or decreased (black asterisks).

In particular, in Fig.~\ref{fig_LargeCouple} we take the first configuration from Fig.~\ref{fig_2BranchNets} to show $\overline{\sigma^2}$'s behaviour around the strong-coupling hysteresis. We highlight that this behaviour in $\overline{\sigma^2}$ (with insignificant variations) appears for all $52$ configurations. In this example, we can see that as we increase [decrease] $\varepsilon$ the critical bifurcation point is $\varepsilon_c^{(i)} \simeq 0.792$ [$\varepsilon_c^{(d)} \simeq 0.776$]. For the remaining $51$ configurations, this hysteresis exhibits minor changes, with slightly different values for the critical bifurcation strengths ($\varepsilon_c^{(i)}$ and $\varepsilon_c^{(d)}$). We can also see that the collapse to the fixed point is also a collapse to a point closer to the synchronisation manifold, since $\overline{\sigma^2} \to 0$ as $\varepsilon \to \epsilon_c^{(i)+}$ (namely, as the coupling is increased beyond its critical value). This can be expected, since for strong coupling a completely synchronous regime is always likely to emerge. However, because our system has parameter differences and noise, a completely synchronous regime is impossible. Hence, as it can be seen, as $\varepsilon$ continues to grow, the fixed point moves away from the (non-existing) synchronisation manifold.
%
\section{Conclusions}
In this work, we present a bifurcation analysis -- with a focus on hysteresis -- for experimentally-coupled logistic circuits, which model a real-system of $6$ nearly identical, chaotic, Kaneko coupled \cite{Kaneko}, logistic maps. The circuit implementation for each logistic map is detailed on our previous work \cite{LHer} (which also shows how closely it approximates to numerical simulations). Here, we extend the circuit to include a coupling block (see Fig.~\ref{fig_CBC}) allowing to choose the coupling configuration -- we explore $52$ different networks of coupled maps. In order to have a tractable framework, we set the system such that the maps are nearly identical, finding that after tuning the experimental resistances, each isolated map responds as having the following parameters: $r_1 = 3.7364$, $r_2 = 3.7537$, $r_3 = 3.7609$, $r_4 = 3.7446$, $r_5 = 3.7298$, and $r_6 = 3.7300$, with a common uncertainty of $3\times10^{-4}$. Moreover, we critically analyse $52$ different coupling configurations, ranging from the ring network to the all-to-all connectivity, which are possible due to our novel coupling circuitry that allows for simple manipulations. Consequently, our experimental bifurcation and hysteresis study is close to those numerical analysis that deal with nearly identical logistic maps, but now, we are also including intrinsic noise and minimal, uncontrolled, parameter mismatch.

In general, we show that this coupled system has robust multi-stable regions with competing attractors, regardless of the coupling configuration. These multi-stable regions are revealed by hysterical regions in the order parameter behaviour, $\overline{\sigma^2}$ (which is the time-average of the quadratic difference between pairs of trajectories, then averaged over all pairs). Specifically, $\overline{\sigma^2}$ holds different values for increasing or decreasing coupling strengths, $\varepsilon$, during the hysterical region. These regions appear in spite of the system's heterogeneity and electronic noise, with particularities depending mainly on the coupling configuration.

Our analyses for the $52$ coupling configurations allowed us to group them according to the $\overline{\sigma^2}$'s behaviour. Overall, we managed to group all the configurations, classifying $47$ configurations under a weak coupling-strength hysteresis region, where we find that $5$ have a single branching [Fig.~\ref{fig_1BranchNets}], $22$ have a double branching [Fig.~\ref{fig_2BranchNets}], $17$ have a triple branching [Fig.~\ref{fig_3BranchNets}], and $3$ having $4$ branches [Fig.~\ref{fig_4BranchNets}] in their $\overline{\sigma^2}$ values as $\varepsilon$ is increased, but a single branch when it is decreased. In spite of this classification, we cannot find a relationship between the underlying connectivity and the particularities of these hysteresis -- which requires further work. On the other hand, we see that all $52$ configurations have a small hysteresis region for strong couplings [Fig.~\ref{fig_LargeCouple}]. This region corresponds to a period-doubling bifurcation at approximately $\varepsilon\simeq 0.8$, where the system transitions from a period $2$ attractor to a fixed point, which is close to the synchronisation manifold.

These analyses also allow us to reveal coupling-strength regions where the collective-dynamics approximates the synchronisation manifold -- which is non-existent when there is parameter mismatch and noise. The reason being that our order parameter, $\overline{\sigma^2}$, effectively measures the trajectory's distance to the synchronisation manifold, which for $6$ identical logistic maps corresponds to the diagonal of the hyper-cube with sides, $[0,\,1]$. In particular, we find that the system is close to the diagonal, i.e., $\overline{\sigma^2}\lesssim 10^{-6}$ for almost all configurations in the region $0.3 \lesssim\varepsilon\lesssim 0.5$, with the exception of the ring-like configurations in Fig.~\ref{fig_RingNets} and the complete-like configurations in Fig.~\ref{fig_CompleteNets}, which remain close to synchronous up to $\varepsilon\lesssim 1.0$. These conclusions could be further corroborated by applying the Master Stability Function approach \cite{Yamada,Pecora} for nearly identical systems \cite{Nishikawa}, which is outside the scope of our current work.

Overall, our results report novel classifications for the emerging multi-stability regions in coupled logistic maps, which are yet to be explained in terms of the connectivity, symmetries, and system size. Moreover, we provide an experimental realisation of a coupled-map system, which is simple to implement, cheap, precise, and reliable -- with a high signal-to-noise ratio ($\sim10^6$).
%
\appendix
%
\section{Circuit details} \label{App_Blocks}
\unskip
    \subsection{Sample-and-hold block}
\begin{figure*}[htbp]
    \centering
    \includegraphics[width=1.9\columnwidth]{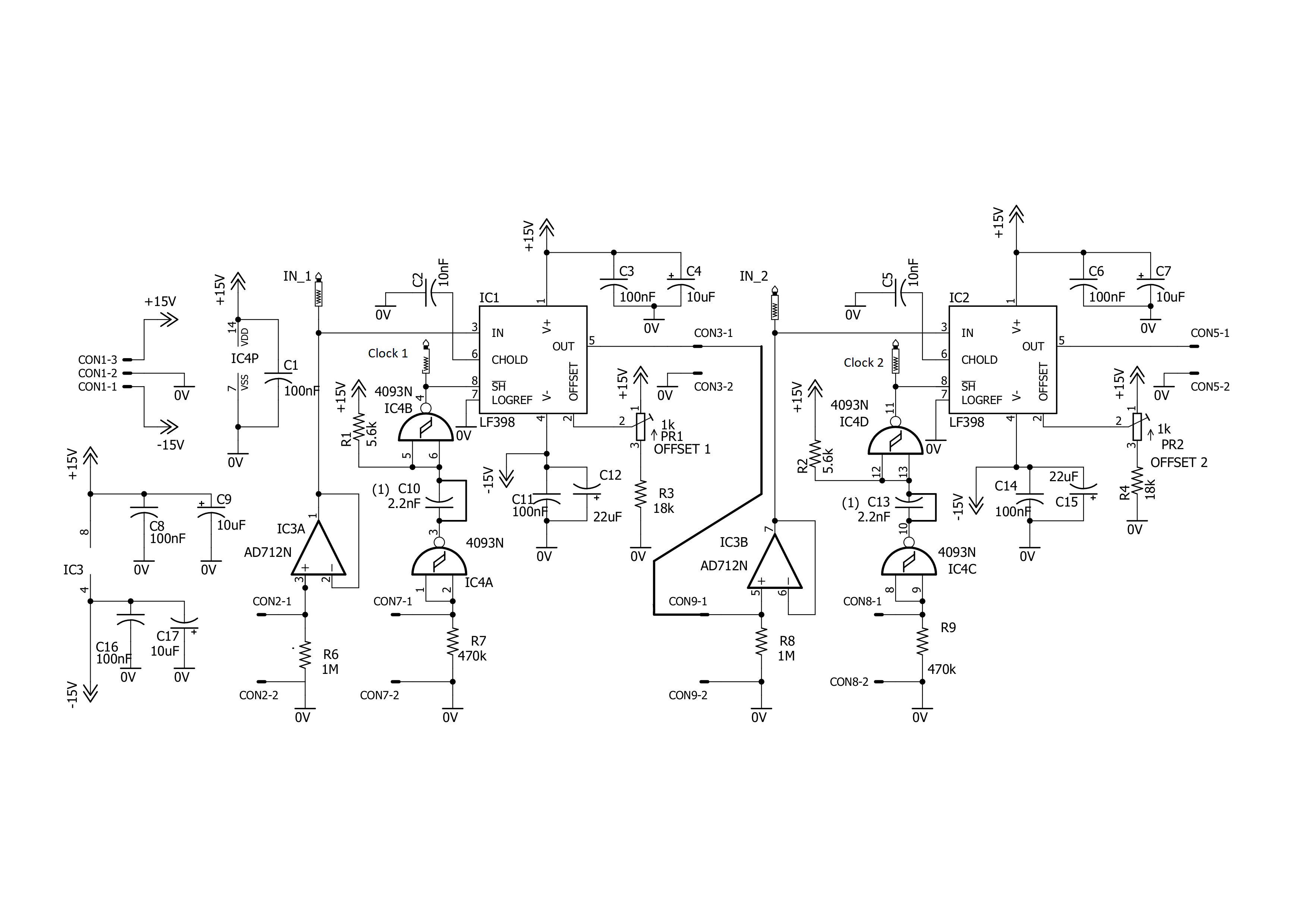}
    \caption{Schematic circuit of the sample-and-hold block (SHB) \cite{LHer}. The SHB is based on $2$ LF398 circuits (the two left-most buffers) and an op-amp circuit (the right-most buffer). A SHB produces a step-wise evolution of the input voltage.}
    \label{fig_SHB}
\end{figure*}

Figure~\ref{fig_SHB} shows the circuit design for the Sample-and-Hold Block (SHB), which is composed of $2$ LF398 circuits and an operational amplifier (op-amp) \cite{LHer}. A clock controls each LF398 times, where after every period these LF398 interchange roles -- one samples whilst the other one holds. Overall, the SHB takes a voltage from its input terminal at an instant of time (sample), keeps it stored in its capacitor (hold), and then releases it at the output terminal one clock-period later. This results in a discontinuous (analog) evolution of the voltage, that allows for the construction of a discrete-time evolution. In particular, the clock period is adjustable, where we set its frequency to $3.5\,k Hz$ (this choice takes into account the trade-off between the speed at which the system stabilises after each periodic switch and the necessity for a fast acquisition of signals).

\unskip
    \subsection{Coupling module}
\begin{figure*}[htbp]
    \centering
    \includegraphics[width=1.9\columnwidth]{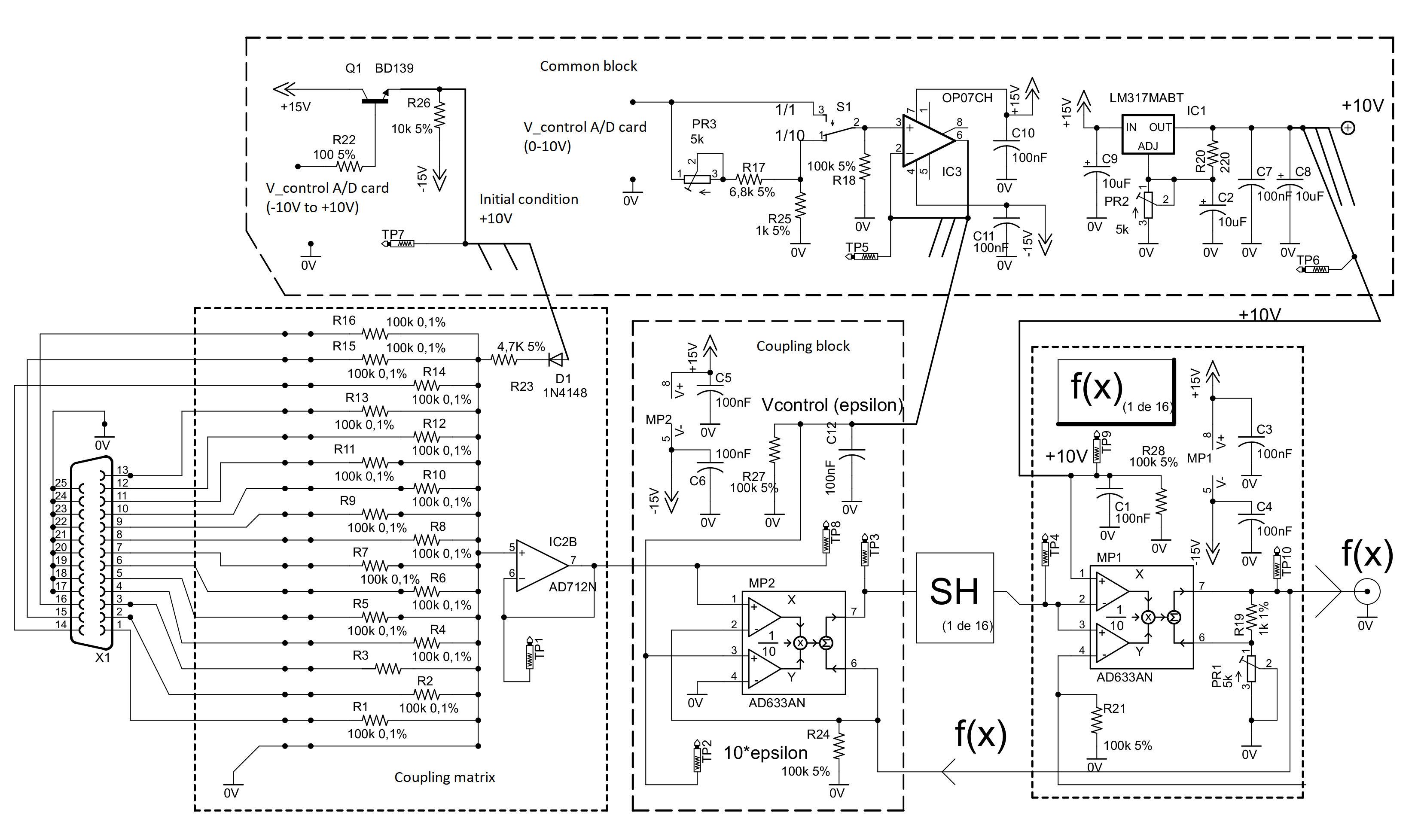}
    \caption{Diagram of the coupling module, which can include $16$ logistic functions and the sample-and-hold block.}
    \label{fig_CBC}
\end{figure*}

Our experimental implementation of the coupling block circuit (CBC) is shown in Fig.~\ref{fig_CBC}. It allows to couple up to $16$ logistic circuits, which are all connected to the SHB to define their discrete-time evolution. Then, they are connected to the CBC to implement the coupling, in accordance to Eq.~\eqref{eq_KanekoLogMaps}.

%
\section{Coupling Configurations} \label{App_Configs}
The $52$ coupling configurations correspond to $52$ adjacency matrices, that we construct by adding [removing] successively a connection from an initial ring network, $\mathcal{C}_{N=6}(k=2)$ [complete network, $\mathcal{C}_{N=6}(k=N-1=5)$], which is a circulant graph with $N = 6$ nodes and degree, $k = 2$ [$k = 5$]. During this process, we discard the networks that are symmetrically equivalent when an interchange of node-labels is carried. Consequently, when adding [removing] $1$ link to $\mathcal{C}_6(2)$ [$\mathcal{C}_6(5)$], we can only define $2$ new networks, which is either adding [removing] a diagonal link or adding a next nearest-neighbour link. When adding [removing] $2$ links to $\mathcal{C}_6(2)$ [$\mathcal{C}_6(5)$], we can only obtain $5$ different network configurations. Finally, we get $9$ different, i.e., non-permutation symmetric, configurations when we add [remove] $3$, $4$, $5$, or $6$ links to $\mathcal{C}_6(2)$ [$\mathcal{C}_6(5)$].
%
\section{Order parameter} \label{App_Param}
The variance between maps $i$ and $j$, $\sigma_{ij}^2$, is given by
\begin{equation}
    \sigma_{ij}^2 = \frac{1}{T}\sum_{n = 1}^T \left( x_n^{(i)} - x_n^{(j)} \right)^2,
    \label{eq_Var2Maps}
\end{equation}
where $\sigma_{ij}^2\to0$ when $x_n^{(i)} \simeq x_n^{(j)}$ for all iterations, $n$ (this happens, for example, if the maps are synchronous), and $\sigma_{ij}^2 = \sigma_{ji}^2$. For our order parameter, we use the system-averaged variance
\begin{equation}
    \overline{\sigma^2} = \frac{2}{N\,(N-1)}\sum_{i = 1}^{N-1}\sum_{j = i+1}^N \sigma_{ij}^2,
    \label{eq_AvgVarMaps}
\end{equation}
where the sums are carried over the ordered pairs of maps [$N(N-1)/2 = 15$].
%
\section*{Acknowledgements}
C.G. acknowledges funds POS\_NAC\_2018\_1\_151237 from the Agencia Nacional de Investigaci{\'o}n e Innovaci{\'o}n (ANII), Uruguay. All authors acknowledge the Comisi{\'o}n Sectorial de Investigaci{\'o}n Cient{\'i}fica (CSIC), Uruguay (group grant ``CSIC2018 - FID13 - grupo ID 722'').

%

\end{document}